\documentclass[sigconf]{acmart}

\usepackage{tabularx}
\usepackage{graphicx}
\usepackage{adjustbox}
\usepackage{color,soul}
\usepackage{xspace}
\usepackage{multirow, booktabs}
\usepackage{makecell}
\usepackage{tabu}
\usepackage{fontawesome5} 
\usepackage{makecell}

\newcommand{\eg}{\textit{e.g.}}
\newcommand{\ie}{\textit{i.e.}}
\newcommand{\etal}{\textit{et al.}}

\AtBeginDocument{%
  \providecommand\BibTeX{{%
    \normalfont B\kern-0.5em{\scshape i\kern-0.25em b}\kern-0.8em\TeX}}}

\setcopyright{acmcopyright}
\copyrightyear{2024}
\acmYear{2024}
\acmDOI{10.1145/3663548.3675635}

\setcopyright{acmlicensed}
\acmConference[ASSETS '24]{The 26th International ACM SIGACCESS Conference on Computers and Accessibility}{October 27--30, 2024}{St. John's, NL, Canada} 
\acmBooktitle{The 26th International ACM SIGACCESS Conference on Computers and Accessibility (ASSETS '24), October 27--30, 2024, St. John's, NL, Canada}

\acmPrice{15.00}
\acmISBN{979-8-4007-0677-6/24/10}

\acmSubmissionID{123-A56-BU3}

\begin{document}

\title{Understanding How Blind Users Handle Object Recognition Errors: Strategies and Challenges}


\author{Jonggi Hong}
\affiliation{%
    \institution{Department of Computer Science}
    \institution{Stevens Institute of Technology}
    \city{Hoboken}
    \state{NJ}
    \country{USA}
}
\email{jhong8@stevens.edu}

\author{Hernisa Kacorri}
\affiliation{%
    \institution{College of Information, UMIACS}
    \institution{University of Maryland, College Park}
    \city{College Park}
    \state{MD}
    \country{USA}
}
\email{hernisa@umd.edu}

\renewcommand{\shortauthors}{Jonggi Hong and Hernisa Kacorri}

\begin{abstract}
  Object recognition technologies hold the potential to support blind and low-vision people in navigating the world around them. However, the gap between benchmark performances and practical usability remains a significant challenge. This paper presents a study aimed at understanding blind users' interaction with object recognition systems for identifying and avoiding errors. Leveraging a pre-existing object recognition system, URCam, fine-tuned for our experiment, we conducted a user study involving 12 blind and low-vision participants. Through in-depth interviews and hands-on error identification tasks, we gained insights into users' experiences, challenges, and strategies for identifying errors in camera-based assistive technologies and object recognition systems.      
  During interviews, many participants preferred independent error review, while expressing apprehension toward misrecognitions. In the error identification task, participants varied viewpoints, backgrounds, and object sizes in their images to avoid and overcome errors. Even after repeating the task, participants identified only half of the errors, and the proportion of errors identified did not significantly differ from their first attempts. Based on these insights, we offer implications for designing accessible interfaces tailored to the needs of blind and low-vision users in identifying object recognition errors. 
\end{abstract}




\maketitle

\section{Introduction}

The field of computer vision has made significant strides, achieving considerable benchmarking rates in object recognition tasks. Yet, despite these advancements, real-world applications often encounter substantial discrepancies between expected and observed performance~\cite{massiceti2023explaining, cao2022s}. Factors such as complex tasks, resource limitations (\eg, mobile device processing), and inputs that deviate from the training data (\eg, classifying images with personal items or cluttered backgrounds collected by a user) pose persistent challenges, leading to higher-than-anticipated error rates in practical scenarios~\cite{alcorn2019strike}. Moreover, the vulnerability of object recognition systems to adversarial attacks further compounds these challenges~\cite{goodfellow2014explaining, kurakin2016adversarial}. While image classifiers hold potential for supporting the blind community in day to day tasks, they are hindered by their inability to effectively convey recognition errors, especially when tactile or olfactory verification is impractical (\eg, distant objects or scenes). Thus, despite advancements, the gap between benchmark performance and real-world usability remains a critical concern for assistive object recognition systems.

In this work, we explore the challenges that blind users face when handling object recognition errors and the strategies they use to overcome them. Specifically, we conduct a user study with 12 blind and low-vision participants, using a two-pronged approach: a semi-structured remote interview and a hands-on error identification task in participants' homes. In the interview, we aim to answer the following research question: \textit{``What are the experiences of blind and low-vision users with error handling in camera-based assistive technologies?''} Participants describe how often they verify recognition results, the frequency with which they encounter errors, the importance they place on these errors, and the challenges they face in identifying them. To better contextualize their responses, we discuss their confidence in photo composition, the frequency of use, and the purposes for each of their camera-based assistive technologies. The interview is then followed by the experiment with an error identification task, where we aim to answer the following research questions: \textit{``How do blind and low-vision users identify and respond to object recognition errors, and what are the relationships between recognition error types, decision-making time, confidence levels, and task repetition?''} Participants interact twice with URCam, an object recognition iOS app that we developed for this experiment. We fine-tuned the underlying model to recognize 15 object stimuli relevant to our study. The app provides object labels or a 'Don’t know' response when the recognition confidence is low. To better contextualize the results, we report the accuracy of URCam during the task and manually code the strategies participants use for capturing photos when URCam responds with 'Don’t know.'

Findings from our study provide insights on blind users' interaction with error-prone object recognition technologies. Interviews indicate that many participants preferred to independently review photo quality and identify errors in camera-based assistive technologies. They often triangulated information using contextual cues, their remaining vision, multiple trials, or other AI apps, rather than seeking sighted assistance. Although the frequency of encountered errors varied among participants, most expressed concern about misrecognitions. However, some did not reported difficulty in identifying these errors. During the error identification task, we observed that participants could identify, on average, only half of the errors, with most of these being false positives.  Notably, participants strategically adjusted viewpoints, backgrounds, and object size to avoid the ``Don't know'' predictions, often rotating objects or the camera to reveal different angles. We found that participants tended to make decisions more quickly when they felt more confident about the accuracy of the predictions. Comparing participants' first and second attempts at the same task, we did not observe a significant difference in the proportion of errors identified. However, there was a notable decrease in time spent to make a decision during the second attempt. Additionally, participants' certainty regarding recognition correctness decreased in subsequent attempts, attributed mainly to inconsistent recognition outcomes among similar objects.

The contributions of this work are the following: 
(1) Providing insights into blind users' experiences in assessing the quality of photos and handling errors in camera-based assistive technology.
(2) Characterizing the challenges encountered by blind people in using object recognition technologies, particularly in error identification and user's confidence.
(3) Suggesting practical implications for the design of object recognition systems, with a focus on error-handling mechanisms, based on empirical findings.

\section{Related Work}
Object recognition, encompassing both object detection and classification~\cite{russakovsky2015imagenet, andreopoulos201350}, has been the subject of active research for decades, representing fundamental and inherently challenging problems within computer vision. Object detection specifically seeks to ascertain the precise location and dimensions of objects within an image, often represented through bounding boxes~\cite{zhao2019object, zou2023object}. On the other hand, image classification aims to determine whether certain objects, belonging to predefined classes, are present within an image or not~\cite{liu2020deep, kuznetsova2020open}. Both object detection and image classification find application in a diverse array of fields, including accessibility. Just within the context of technologies for blind and low vision people, the focus of this paper, there are a myriad of publications. In a recent review by Gamage \etal, the breakdown highlights the various assistive tasks where this technology is being utilized, covering a wide range of contexts from \textit{handling object and devices},  \textit{orientation and mobility}, \textit{communication and information}, \textit{personal care and protection}, \textit{cultural and sports activities}, to \textit{personal medical treatment}~\cite{gamage2023what}. Given the inherently error-prone nature of this technology, understanding and designing for user interactions with prediction errors is critical. Below we synthesize prior literature that discuss this in the context of assistive technologies for the blind and more broadly.

\subsection{Interactions with Errors in AI Technology in the Context of the Blind Community}

\begin{table}
\small
\centering
\caption{Characteristics of related studies on errors in AI-infused assistive technology juxtaposed with ours.}
        \setlength\tabcolsep{3pt}
       \begin{tabular}{l r c c c c c c c c c c}
            \toprule
              & & \cite{macleod2017understanding} &\cite{salisbury2017toward} & \cite{brewer2018understanding} & \cite{saha2019closing} & \cite{hong2020reviewing} & \cite{alharbi2022understanding}& Ours \\
            \midrule
            \multirow{2}{*}{\textbf{\rotatebox[origin=c]{90}{People}}}
            & Blind \& low vision    & 6, 100    & 7         & 15        & 22, 13    & 12        & 20        & 12        \\
            & Sighted         &           & 235       &           &           & 12        &           &           \\ [0.2cm]            
            \midrule
            \multirow{2}{*}{\textbf{\rotatebox[origin=c]{90}{Input}}}
            & Photo                 & $\bullet$ & $\bullet$ &           &           &           & $\bullet$ & $\bullet$ \\
            & Speech                &           &           &           &           & $\bullet$ &           &           \\
            & Other                 &           &           & $\bullet$ & $\bullet$ & $\bullet$ &           &           \\
            \midrule
            \multirow{4}{*}{\textbf{\rotatebox[origin=c]{90}{Methods}}}
            & Interview             & $\bullet$ &           &           & $\bullet$ & $\bullet$ & $\bullet$ & $\bullet$ \\
            & Survey                &           & $\bullet$ &           & $\bullet$ &           &           &           \\
            & Focus group           &           &           & $\bullet$ &           &           &           &           \\
            & Crowdsourcing         &           & $\bullet$ &           &           &           &           &           \\
            & Lab study   & $\bullet$ &           &           &           & $\bullet$ &           & $\bullet$ \\
            \midrule
            \multirow{3}{*}{\textbf{\rotatebox[origin=c]{90}{Task}}}
            & Image captioning      & $\bullet$ & $\bullet$ &           &           &           &           &           \\
            & Speech recognition    &           &           &           &           & $\bullet$ &           &           \\
            & Object recognition    &           &           &           &           &           &           & $\bullet$ \\
            & Navigation            &           &           &           & $\bullet$ &           &           &           \\
            & Obfuscation           &           &           &           &           &           & $\bullet$ &           \\
            & Controlling a car     &           &           & $\bullet$ &           &           &           &           \\
        \bottomrule
    \end{tabular}%
\label{tab:related_work}
\end{table}

Previous research has consistently demonstrated the significant impact of errors on the experiences of blind users. Table~\ref{tab:related_work} illustrates previous research examples concerning the ramifications of errors in AI-infused assistive technology. For instance, safety concerns regarding malfunctions in autopilot systems of self-driving vehicles pose a primary apprehension for blind individuals who are encouraged to use such vehicles autonomously~\cite{brewer2018understanding, brinkley2017opinions}. Similar concerns arise in systems where error risks are less critical than those in self-driving vehicles but still consequential. 
For instance, studies have shown that minor errors in navigation systems can lead to frustration and disorientation, even when the destination is just a few meters away (\eg, \cite{saha2019closing}). Prior studies also highlighted the need for blind users to distinguish and handle errors when understanding images with AI-based image descriptions~\cite{lee2022imageexplorer, gonzalez2024investigating}. These findings underscore the importance of user-error interaction interfaces that provide contextual information and predictions from machine learning models to help blind users accurately assess error causes and severity.

Similarly, errors significantly impact blind users' experiences with object recognition systems, as blind individuals often rely solely on system outputs due to the challenge of verifying them~\cite{morris2020ai, salisbury2017toward}. Consequently, understanding the implications of errors in AI-infused assistive technology is critical. Research has highlighted instances where such errors have led to adverse outcomes. For example, blind users tend to overtrust automatically generated captions on social media images, even when the captions are incorrect and nonsensical~\cite{macleod2017understanding}. While some errors in blind navigation systems are manageable in familiar environments, they become problematic when they can lead to embarrassing situations with bystanders~\cite{abdolrahmani2017embracing, lee2020pedestrian}. Moreover, errors in image recognition systems used for controlling household objects can pose safety threats. Consequently, robust safety mechanisms are essential for such tools. 
Given the significance of error handling in object recognition systems for blind users, this work delves into and delineates the challenges they face in identifying and recovering from object recognition errors with the number of participants, methods, and task contextualized within this literature.

\subsection{Interactions with Errors in AI-infused Technology in a Broader Context}
While errors are easy to tell in some applications where users can understand the outcome from the system and ground truth easily (\eg, navigating familiar routes with a way-finding system), the outcome from the system may not be clearly perceived due to the characteristics of the task, a poorly designed interface, the complexity of the information, or poor concentration caused by a high workload~\cite{kontogiannis1999user, kontogiannis2009proactive}. For example, the ground truth may not be available immediately when the outcome is provided by the system (\eg, medical diagnosis, weather prediction). The ground truth may not be straightforward to the user if the system handles data in an unfamiliar work domain~\cite{sellen1994detection}. Therefore, many researchers have worked on developing user interfaces for AI-infused systems aimed at effectively managing errors and aiding users in navigating discrepancies between system outputs and desired outcomes. Noteworthy efforts include strategies to temper user expectations regarding AI system performance~\cite{kocielnik2019will, lukashova2023influence}, alongside the presentation of user-friendly interfaces tailored to address errors arising from diverse AI-infused applications.

Gesture recognition technology has found utility in controlling an array of devices, from visual displays~\cite{lee2020development} and robots~\cite{meghana2020hand} to wearable devices~\cite{nooruddin2020hgr, tan2022self}, and even in virtual reality interactions~\cite{gupta2020hand, huang2021evaluation, chen2021gestonhmd}. Despite considerable advancements in gesture recognition accuracy and usability, input recognition errors persist, significantly detracting from user experiences~\cite{lafreniere2021false}. Research endeavors have thus delved into comprehending the ramifications of these errors, uncovering that user tolerance is frequently shaped more by the context of interaction than solely by system performance. Remarkably, users may tolerate recognition error rates of up to 40\% before opting for alternative interaction modes over gestures~\cite{karam2006investigating}. Moreover, endeavors to alleviate the detrimental impacts of gesture recognition errors have encompassed various strategies, such as real-time error detection and adaptive model adjustments based on discerning whether the erroneous inputs stem from user mistakes or recognition errors~\cite{sendhilnathan2022detecting}.

Similarly, in the context of speech recognition systems, while significant progress has been made in minimizing errors under controlled environments, practical challenges such as speaker variability and ambient noise persist~\cite{goldwater2010words, jiang2005confidence, myers2018patterns}. These errors manifest in various forms, including failure to detect speech, misrecognition, or incorrect handling of recognized speech~\cite{hong2020reviewing, pearl2016designing}. Studies have revealed that users overlook more than half of speech recognition errors in the absence of visual cues~\cite{hong2020reviewing}. To address this, researchers have explored techniques for automated error detection in speech recognition outputs, ranging from visually highlighting potentially erroneous words to employing neural network-based predictive models~\cite{errattahi2018automatic, berke2017deaf, tam2014asr, ghannay2015asr, ghannay2015word}. However, despite advancements, predicting speech recognition errors remains an ongoing research area, with current methods achieving moderate precision and recall rates.

Beyond gesture and speech, researchers are also endeavoring to mitigate errors in other AI-infused applications, including robotics~\cite{liu2023robot} and autonomous vehicles~\cite{wang2022and}. In these domains, where safety and reliability are paramount, error-handling mechanisms play a critical role in ensuring smooth operation and user trust~\cite{ajenaghughrure2020risk, raats2020trusting}. Strategies such as fault tolerance, redundancy, and fail-safe mechanisms are being explored to minimize the impact of errors and safeguard against catastrophic failures~\cite{macrae2022learning, athavale2020ai, wu2022fault}. Moreover, advancements in simulation and testing methodologies enable researchers to systematically evaluate robustness and user experience with errors in real-world deployment scenarios~\cite{afzal2020study, afzal2021simulation, perello2021driver}. Overall, the quest for error-resilient AI-infused systems represents a multifaceted and interdisciplinary endeavor, requiring collaboration across domains to achieve the vision of intelligent, trustworthy technology.

\section{Methods}
To gain insight into blind people's challenges and strategies in handling errors in
AI-infused applications for object recognition, we carry out a comprehensive two-phase user study. The study first encompasses a semi-structured interview that captures participants' experience with camera-based assistive tools. The interview is then followed by an object recognition task, where participants are asked to identify errors when interacting with a mobile application in their homes. We adopt this two-pronged approach from a prior study by Hong \etal~\cite{hong2020reviewing} looking at challenges and strategies adopted by blind people when reviewing automatic speech recognition errors. Our study was approved by the Institutional Review Board at our \textit{anonymized institution} (IRB \textit{number anonymized}). Participants were compensated at a 15\$/hour rate for a total of \$26.21 on average ($\$23-29, SD=1.73$). 

\subsection{Participants}
We recruited 12 blind participants (6 women, 6 men, 0 nonbinary) from campus email lists and local organizations. As shown in Table~\ref{tab:participants}, their age ranged from 32 to 70 ($M = 54.3, SD = 15.2$).  Three participants reported being totally blind, five having some light perception, and four being legally blind.   P1 and P2 reported ``\textit{an auditory processing disorder}'' and difficulty hearing ``\textit{very high sounds}'', respectively. Yet, all participants indicated that they faced no problems in using a screen reader.  All mentioned using smartphones several times a day. All participants were right-handed except for one, who was left-handed (P4).
When asked to report their levels of familiarity with machine learning, two participants reported being somewhat familiar, eight being slightly familiar, and two being not familiar at all. We used a 4-point scale for this question, where \textit{not familiar at all} indicated that participants have never heard of machine learning, \textit{slightly familiar} that they have heard of it but don’t know what it does, \textit{somewhat familiar} that they have a broad understanding of what it is and what it does, and \textit{extremely familiar} that they have extensive knowledge on machine learning. All questions are available in Appendix~\ref{appendix_interview_questions}.

\begin{table}[t]
    \small
    \centering
    \begin{tabu}{l l l l l l} 
        \hline
        ID  & Age & Gender & Level of vision  & Onset& Familiarity with ML*\\
        \hline
        P1  & 39  & Female & Light perception & Birth & Not familiar at all \\
        P2  & 67  & Male   & Legally blind    & 55    & Slightly familiar \\
        P3  & 62  & Female & Totally blind    & Birth & Somewhat familiar \\
        P4  & 32  & Male   & Legally blind    & 20    & Slightly familiar \\
        P5  & 66  & Male   & Light perception & 46    & Slightly familiar \\
        P6  & 61  & Male   & Light perception & 41    & Somewhat familiar \\
        P7  & 70  & Male   & Legally blind    & Birth & Slightly familiar \\
        P8  & 50  & Female & Legally blind    & 45    & Slightly familiar \\
        P9  & 69  & Female & Totally blind    & 55    & Not familiar at all \\
        P10 & 66  & Female & Light perception & Birth & Slightly familiar \\
        P11 & 33  & Female & Light perception & Birth & Slightly familiar \\
        P12 & 36  & Male   & Totally blind    & Birth & Slightly familiar \\
        \hline
    \end{tabu} 
    {\small *ML: Machine learning}
\caption{Participants' demographics and background.}
\label{tab:participants}
\end{table}


\subsection{Procedure}

The study is conducted over two days that may be up to 7 days apart. On the first day,  participants engage in a semi-structured interview and answer questions related to demographics, and technology experience.  On the second day, they complete a recognition task with an object recognition application engineered by our team that aims to serve as a testbed. The app is called URCam. During this session, participants interact with URCam and a set of given object stimuli. They attempt to identify any recognition errors that the app might have made and express their confidence.  

\subsubsection{Semi-structured interview.} The interview lasted 51 minutes on average ($18-90m, SD=21.37$). It was completed remotely over Zoom and recorded for later analysis.  Beyond demographics, participants responded to questions about:
\begin{itemize}
    \item frequency of using a mobile device, taking photos, reviewing photos, and changing settings of the camera;
    \item purpose of taking photos, subjects included, applications and devices used, and confidence on photo composition;
    \item frequency of use of a camera-based assistive application, its usefulness, and device;
    \item frequency of verifying the recognition results of a camera-based assistive application, encountering errors, importance of errors, and difficulty of identifying the errors; 
    \item strategy of taking photos with an assistive application, degree of understanding how that application works.
\end{itemize}

As shown in Appendix~\ref{appendix_interview_questions}, questions assessing frequency are categorized into two groups. The first includes those answerable with an absolute 7-point scale, adopted from Rosen \etal~\cite{rosen2013media} (ranging from `\textit{never}' to `\textit{several times a day}'). For example,  `\textit{How often do you take photos or record a video?}'
The second group includes those suited to a relative 6-point scale (from `never' to `always')~\cite{brown2010likert} \eg, `\textit{How often do you encounter misrecognitions when you use Seeing AI?}'

\subsubsection{Error identification task.} 
\begin{figure}[t]
    \centering
    \includegraphics[width=1\columnwidth]{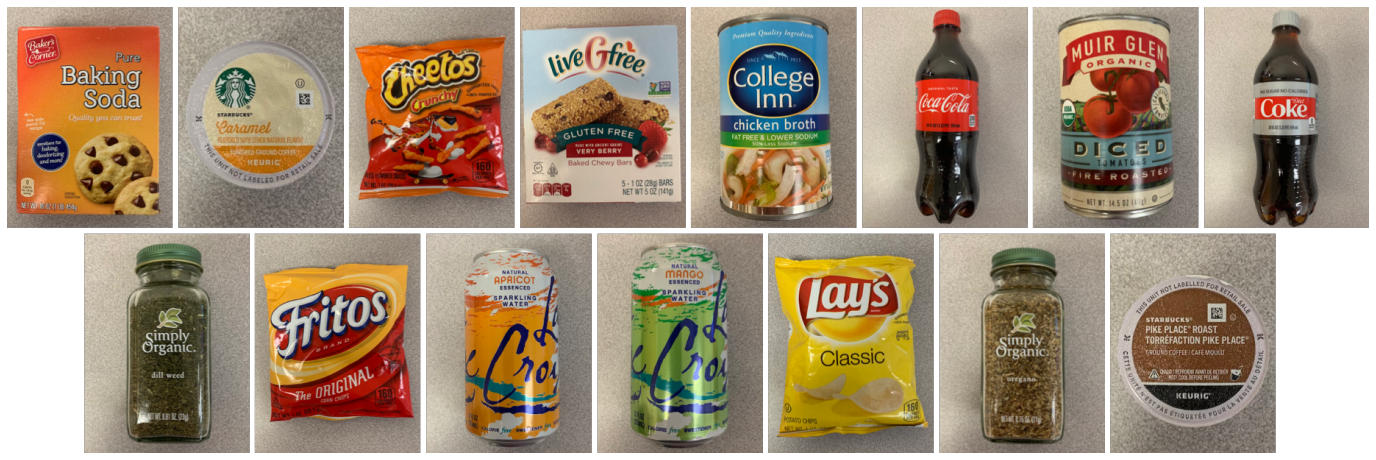}
    \caption{Object stimuli in our study from Kacorri \etal~\cite{kacorri2017people}: baking soda, caramel coffee, Cheetos, chewy bars, chicken broth, coca-cola, diced tomatoes, diet coke, dill, Fritos, Lacroix apricot, Lacroix mango, Lays, oregano, roast coffee.}
    \label{fig:tor_study_urcam_objects}
    \Description{The figure shows 15 photos showing the objects in the caption.}
\end{figure}

Given a set of object stimuli and an iPhone 8 device with an object recognition app, participants are asked to try to identify the objects using the application. When deployed in real-world environments, object recognition errors are typically confounded by blurred images, viewpoints with low discriminative characteristics, cluttered backgrounds, low saliency, and more importantly partially included or out-of-frame objects of interest~\cite{lee2019hands, chiu2020assessing, bafghi2023new}. Thus, we do not conduct this session in our lab, but move the study to the homes of blind participants. As in Lee \etal~\cite{lee2022lab}, all study materials are delivered at home, and instructions are conveyed via Zoom. Each participant received a laptop where the Zoom call is set up for remote communication. Furthermore, participants are provided with Vuzix Blade smart glasses, featuring an integrated camera and initiated with the Zoom call. The smart-glasses can both enable real-time access to participants' first-person perspectives and allow for recordings of observations for subsequent data analysis.  At the beginning of the task, the experimenter presents a list of 15 objects for reference (Figure~\ref{fig:tor_study_urcam_objects}). During each trial, participants randomly select an object, capture its image, and obtain a label from the object recognition app, which is communicated via synthesized speech. Upon hearing ``\textit{Don't know}'' from the app, indicating that it failed to recognize any object in the photo, participants proceed to capture additional photos until the app provides a label for an object. Subsequently, participants indicate whether the recognition was accurate and express their confidence level in their judgment of correctness for the recognition. After completing the initial 15 trials with all objects (\textit{Attempt 1}), participants repeat the process with the objects in a randomized order (\textit{Attempt 2}), totaling 30 trials. Participants are encouraged to think aloud throughout the task. Upon task completion, participants provide feedback on the difficulty level and the strategies they employ for identifying errors.

\subsection{Object Stimuli}
For the error identification task, we utilize a fixed set of 15 objects across all participants (Figure~\ref{fig:tor_study_urcam_objects}). We adopt similar stimuli to those previously employed in a study examining the interaction of blind users with a teachable object recognizer by Kacorri \etal~\cite{kacorri2017people}. We adopt their methodology, which involved the selection of objects to encompass a variety of shapes, sizes, materials, and visual similarities. While some products, such as baking soda, chicken broth, diced tomatoes, and diet coke, featured logos or images on their containers that differed slightly from those used in the prior study due to design updates, the fundamental aspects affecting participants' tactile perception, such as shape, material, and weight, remained consistent across all objects.

\subsection{URCam: An Object Recognition App}
\begin{figure}[t]
    \centering
    \includegraphics[width=0.7\columnwidth]{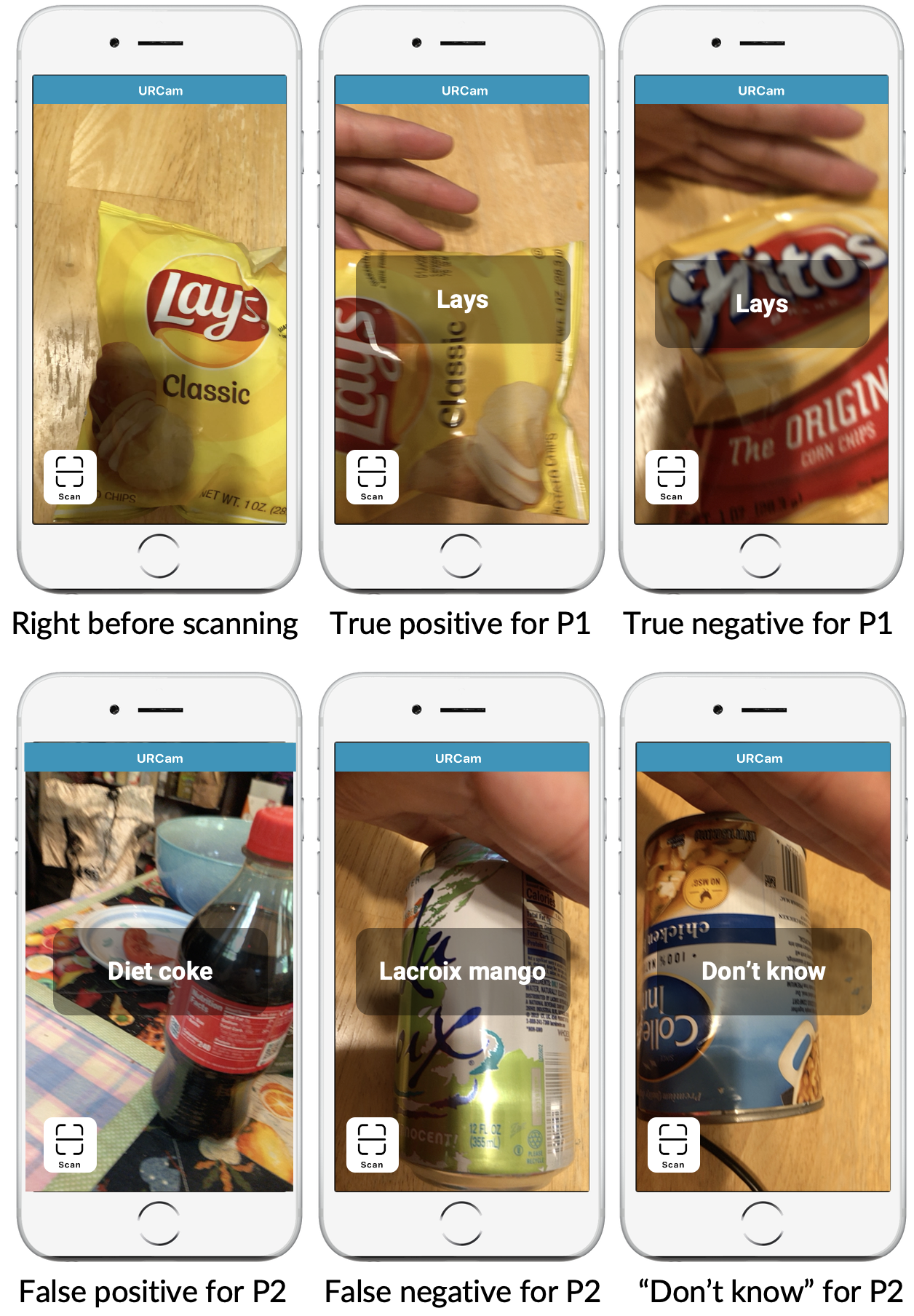}
    \caption{A series of screenshots from URCam that was deployed in the study, where participants P1 and P2 experienced correct, incorrect, and uncertain predictions communicated via a ``Don't know'' message.}
    \label{fig:tor_study_urcam}
    \Description{A series of screenshots from an app named URCam displayed on a smartphone. The first screen shows a photo of Lays chips and Lays as a recognition result with the text 'Right before scanning'. The second screen shows Lays chips again and Lays as a recognition result with the text 'True positive for P1'. The third screen displays Fritos chips and Lays as a recognition result with the text 'True negative for P1'. The fourth screen presents Coca cola image and Diet Coke as a recognition result with the text 'False positive for P2'. The fifth screen exhibits LaCroix mango-flavored sparkling water image and Lacroix mango as a recognition result with the text 'False negative for P2'. Lastly, the sixth screen showcases a chicken broth can item labeled 'Don't know' with the text 'Don’t know for P2'.}
\end{figure}

For the error identification task, we build an object recognition app, called URCam, that serves as a testbed; the software used as a basis for experimentation. URCam is fine-tuned using the images of objects in Figure~\ref{fig:tor_study_urcam_objects}. The base model of the object recognizer is InceptionV3~\cite{szegedy2016rethinking}, originally trained on the ImageNet dataset~\cite{deng2009imagenet}. The dataset for fine-tuning comprises photos captured by nine blind participants in a previous study by Lee \etal~\cite{lee2019revisiting}, where they trained a teachable object recognizer. Their dataset includes 225 images for each object, totaling 3375 images. Although other existing datasets (\eg, \cite{bafghi2023new}) provide images collected by blind and low-vision people, they did not include fine-grained labels for the specific objects in our study. Therefore, we opted for the dataset collected with blind participants that included those objects. Fine-tuning involves 500 iterations of gradient descent with a learning rate of 0.01. During the identification task, our study participants interact with URCam on an Apple iPhone 8. As shown in Figure~\ref{fig:tor_study_urcam}, upon pressing the \faMinusSquare[regular]~\textit{Scan} item button, the app transmits the image to a server via HTTP, where the fine-tuned object recognition model generates predictions regarding the image's label, subsequently relaying it back to the participant's device via voice and visual display.

To differentiate between objects within our training set and those the app hasn't encountered previously, we employ a technique that assesses the model's discriminative capacity by measuring the entropy of its confidence scores~\cite{zhang2012online}. Specifically, we establish a threshold for both the entropy value and the confidence score to determine instances where the model should refrain from providing a predicted label and instead output ``\textit{Don't know}''. If the entropy value exceeds 2.0 or the confidence score falls below 0.4, the application synthesizes the phrase ``\textit{Don't know}'' rather than presenting a predicted label. With this precautionary measure, the model strives to abstain from delivering potentially misleading or inaccurate predictions when it lacks sufficient confidence in its discriminatory abilities.

\subsection{Data Analysis}
The responses from the semi-structured interview and tasks are captured via Zoom. We transcribe these responses to enable a comprehensive analysis of the participants' experience and feedback. We also explore how participants handle application uncertainty (\ie, ``\textit{Don't know}'') and deal with potential misrecognitions during the error identification task.

\subsubsection{Semi-Structured Interview.}  We use a thematic coding approach to find the major themes in the participants’ responses~\cite{braun2006using}.  To reduce the subjectivity, two researchers cooperate to code the responses.  One of the researchers transcribes the responses.  With the transcribed data, the two researchers code the responses independently and create initial codebooks.  They compare the two codebooks and code data to resolve the disagreements through consensus.  After resolving the disagreements (a total of 35 out of 373 answers), they establish a shared codebook and code the data.  In the final codebook, the responses of 17 open questions in the semi-structured interview include a total of 153 codes.

\subsubsection{Error Identification Task.}  
We manually annotate the images captured by participants and compare these annotations with the object recognition results recorded by the app to assess the accuracy of object recognition during the task. We categorize the trials based on how well participants identify any object recognition errors by analyzing their responses captured in the video recordings. 
During this analysis, if participants cannot tell whether the recognition was correct or incorrect, which happened for a total of 7 trials, we interpret this as them perceiving that the recognition can be incorrect but being very uncertain about it. Specifically, we group the trials into:

\begin{description}
    \item[True positive:] The object recognition is \textit{correct} and the participant perceive it as \textit{correct}.
    \item[False positive:] The object recognition is \textit{incorrect}, but the participant perceive it as \textit{correct}.
    \item[True negative:] The object recognition is \textit{incorrect} and the participant perceive it as \textit{incorrect}.
    \item[False negative:] The object recognition is \textit{correct}, but the participant perceive it as \textit{incorrect}.
\end{description}

We examine the correlation between participants' confidence levels, and trial completion time, along these 4 groups. Trial completion time is manually measured through video analysis, which involves recording the elapsed time from when the app provided the recognition result to when the participant reports its correctness to the experimenter. Additionally, we investigate any adjustments in participants' strategies for capturing photos when they receive a ``\textit{Don't know}'' response from the URCam. We categorize the adjustments by looking at variation in \textit{background}, \textit{viewpoint}, \textit{illumination}, and \textit{object size}; a coding scheme adopted by Hong~\etal~\cite{hong2020crowdsourcing}.

\section{Insights from the Interview}
The central themes explored during the interview encompass blind people's experiences with photography or video recording and their interaction with camera-based assistive applications. Our discussion delve into various aspects, such as how blind people assess the quality of their photographs, the motivations behind their photography, and the methods they employ to discern inaccuracies within camera-based assistive apps.

\subsection{Capturing and Reviewing Photos}
\begin{figure}[t]
    \centering
    \includegraphics[width=1\columnwidth]{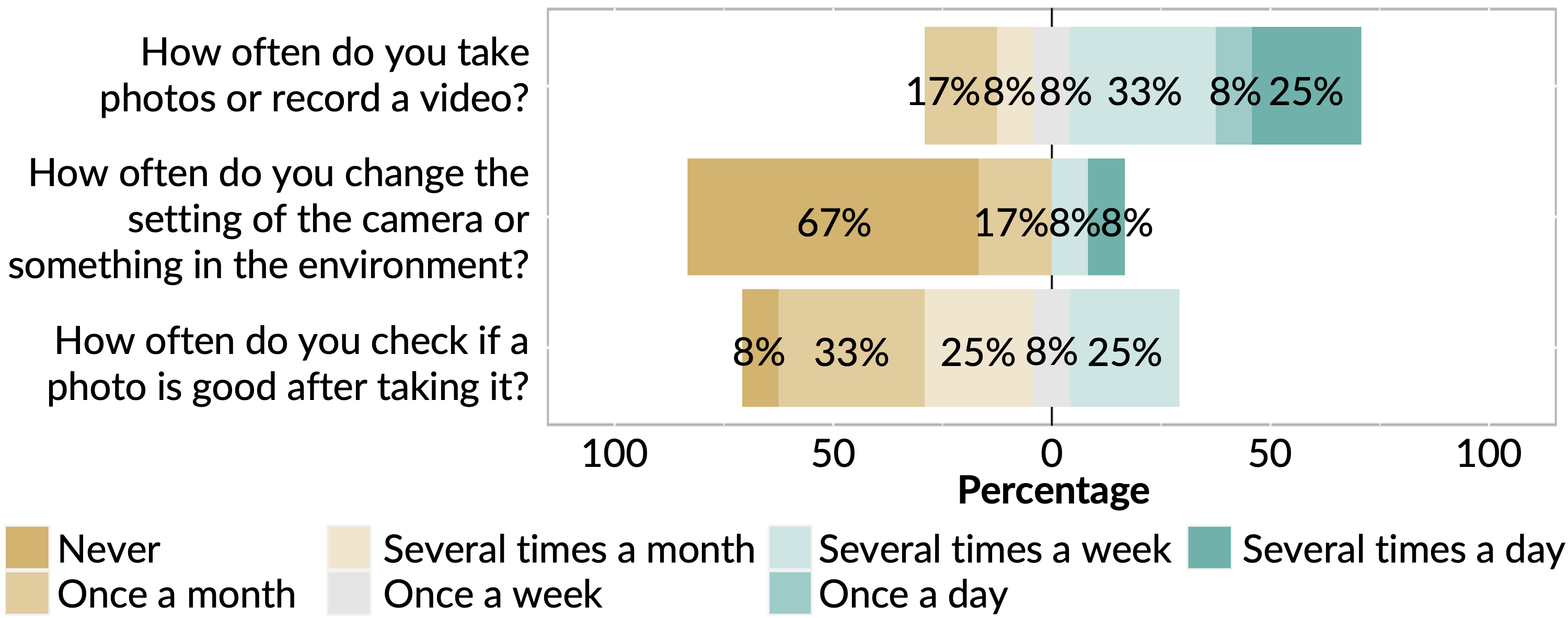}
    \caption{Participants' experience in taking photos.}
    \label{fig:tor1-interview-phototaking-subjective}
    \Description{A horizontal grouped bar chart displaying responses to three questions about photography habits. The questions are: 'How often do you take photos or record a video?', 'How often do you change the setting of the camera or something in the environment?', and 'How often do you check if a photo is good after taking it?'. The x-axis represents the percentage of respondents, ranging from 0\% to 100\%. The y-axis displays the frequency categories: 'Never', 'Once a month', 'Once a week', 'Several times a week', 'Several times a day'. The bars are color-coded to represent different response frequencies. The values displayed on the bars are as follows:

    'How often do you take photos or record a video?':
    Never: 67\%
    Once a month: 17\%
    Once a week: 8\%
    Several times a month: 8\%
    Several times a week: 0\%
    Several times a day: 0\%
    'How often do you change the setting of the camera or something in the environment?':
    Never: 33\%
    Once a month: 17\%
    Once a week: 25\%
    Several times a month: 8\%
    Several times a week: 8\%
    Several times a day: 8\%
    'How often do you check if a photo is good after taking it?':
    Never: 25\%
    Once a month: 8\%
    Once a week: 25\%
    Several times a month: 0\%
    Several times a week: 33\%
    Several times a day: 8\%}
\end{figure}

By delving into participants' experiences with capturing photos or videos, our goal is to uncover the degree of integration of these technologies into the daily routines of blind people. Furthermore, through an exploration of the techniques participants employed to manipulate camera settings, we aim to uncover insights into their approach for capturing photos that would allow them to achieve their goal be it sharing them with others or completing visual tasks. We find that all participants consistently engage in photography activities, each capturing photos at least once a month, as depicted in Figure~\ref{fig:tor1-interview-phototaking-subjective}. This aligns with findings from a previous study indicating that BLV people actively use cameras for daily tasks~\cite{jayant2011supporting}. One of the primary reasons for using a camera was to share images or videos via social media or video calls as shown in prior studies~
\cite{jayant2011supporting, seo2021understanding}. The majority (8 out of 12) report taking photos or videos more frequently than several times a week. One reason for using a camera was to share photos or engage in video calls. For instance, P4 explained, \textit{``Video calls, share photos, I take videos of bands as I play songs. I've got a YouTube channel with several hundred videos of shows I've gone to.''} Additionally, using assistive technology was cited as another reason, as described by P9: \textit{``Sometimes I'll check to see what SeeingAI will say. Just curious to know, what the app will say about. I've used glasses with an app Aira. [...] if you're in like Walgreens, you can connect with Aira and they will tell you what's on the shelf.''} During their photographic endeavors, participants tend to maintain consistent camera settings and environmental conditions. The majority (8 out of 12) of participants reported never altering their camera settings. Among the participants who did make adjustments (4 out of 12), modifications primarily aimed to optimize lighting conditions. Specifically, three participants sought out locations with ample natural light, while one experimented with flash settings. For instance, P7 sought to evade shadows, stating, \textit{``I'll strategically reposition them to ensure optimal lighting without an excess of shadows or other visual distractions.''} Additionally, one participant (P8) explored varying camera angles to enhance their photographic outcomes. P8 expressed a preference for home photography due to the favorable lighting conditions and exploring camera angles, explaining, \textit{``when I'm home, I feel it gives me the maximum amount of light and I get the best pictures. [...]  I might move it around a couple of times so that it'll describe it in the most detailed way.''} 

We also posed questions regarding how often they review their photos, as this practice may influence the quality of their images. In general, participants did not frequently review their photos. The majority (8 out of 12) reported checking their photos several times a month or less. \textbf{Most participants reviewed their photos independently without sighted help.} Participants who identified as legally blind ($N=4$) predominantly relied on their own visual assessment. They utilized automatically generated image descriptions from assistive tools like Seeing AI and iOS's built-in image captioning function ($N=5$). For instance, P12 would judge the quality of their photo based on text recognition results, stating, \textit{``What's relevant are the OCR results I get from it. Especially if there is a garbled section that doesn't fall into a normal OCR error pattern, then I know the photo's not good.''} A possible reason for independent reviewing behavior could be concerns about privacy issues when sharing their photos with sighted people~\cite{xie2024bubblecam, akter2020uncomfortable, stangl2023dump, zhang2023imageally}. Few (3 out of 12) participants sought assistance from sighted individuals in their vicinity and only one (P1) utilized remote assistance through apps like Aira~\cite{Aira} and BeMyEyes~\cite{BeMyEyes}.

To provide context for understanding the motivations behind participants' photography, we asked questions about the subjects they captured in their photos. Participants cited documents for text recognition ($N=10$), people ($N=9$), objects ($N=8$), food ($N=6$), landscapes ($N=5$), and miscellaneous items such as a scene and a bill ($N=4$). Similarly, the most prevalent purposes for taking photos or recording videos were for text recognition ($N=10$), video calls ($N=8$), and object recognition ($N=5$). These responses diverge somewhat from the findings of a previous study conducted by Jayant \etal~\cite{jayant2011supporting} in 2011, which suggested that blind individuals primarily took photos to capture friends or family for leisure, while their most sought-after camera function was text recognition. \textbf{This result indicates the increasing prevalence of computer vision-based assistive applications among blind and low-vision people.} However, many participants still found image framing challenging ($N=9$), a difficulty highlighted in prior studies~\cite{jayant2011supporting, lee2019revisiting, ahmetovic2020recog}. For instance, P1 and P5 expressed concerns such as, \textit{``Making sure the information I'm trying to capture is in the frame of the camera,''} and \textit{``I don't know how far away from the object to hold the phone''}, respectively. Participants also identified other difficulties such as maintaining focus on the object ($N=2$), stabilizing the camera ($N=2$), adjusting lighting conditions ($N=2$), and orienting objects correctly ($N=2$).

\subsection{Handling Image Recognition Errors}
\begin{figure}[t]
    \centering
    \includegraphics[width=1\columnwidth]{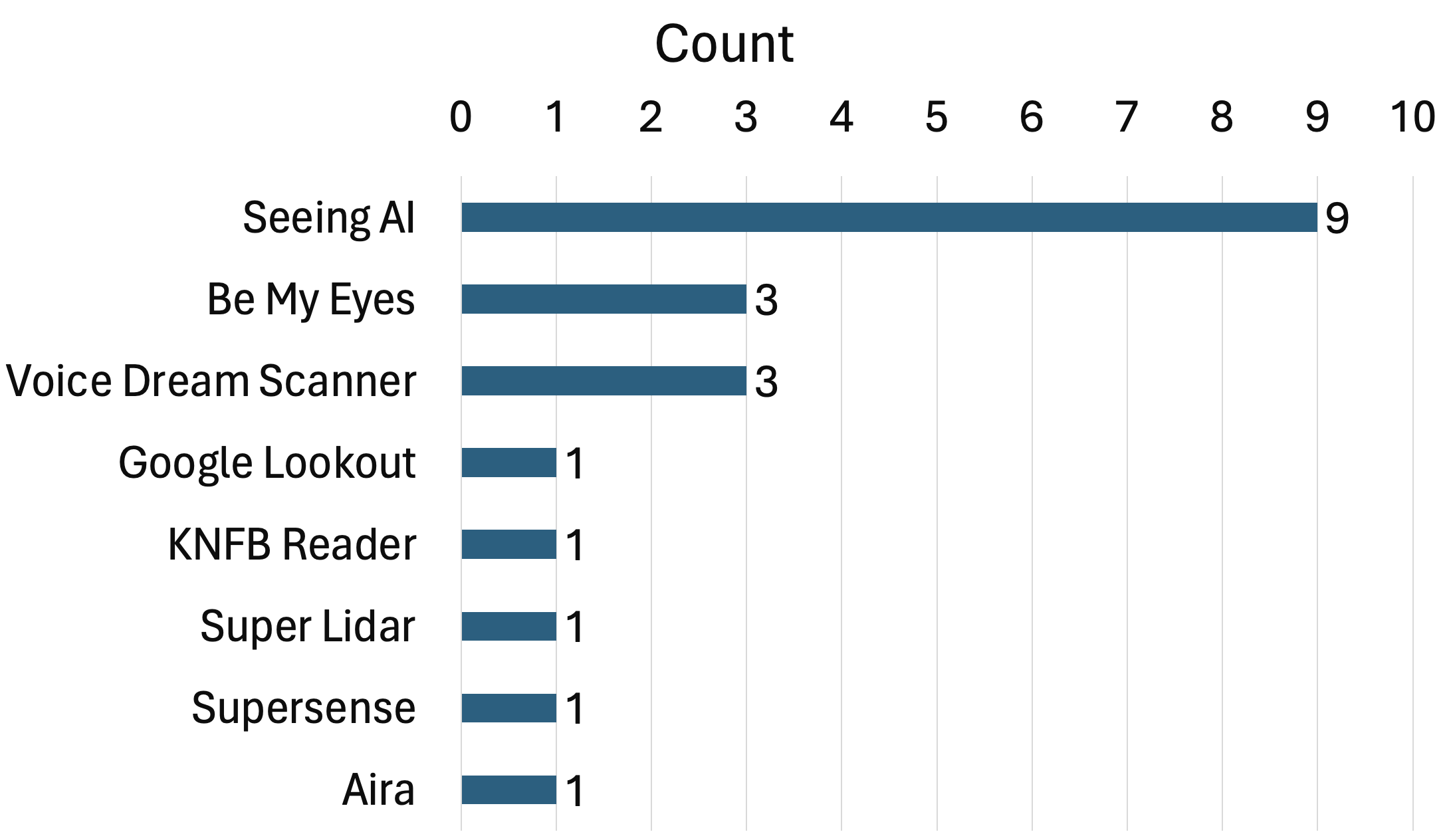}
    \caption{Camera-based assistive apps the participants have used regularly.}
    \label{fig:tor1-interview-apps}
    \Description{A horizontal bar chart titled 'Count' displaying the count of different assistive technology apps mentioned. The apps and their respective counts are as follows: Seeing AI (9), Be My Eyes (3), Voice Dream Scanner (3), Google Lookout (1), KNFB Reader (1), Super Lidar (1), Supersense (1), Aira (1).}
\end{figure}

\begin{figure}[t]
    \centering
    \includegraphics[width=1\columnwidth]{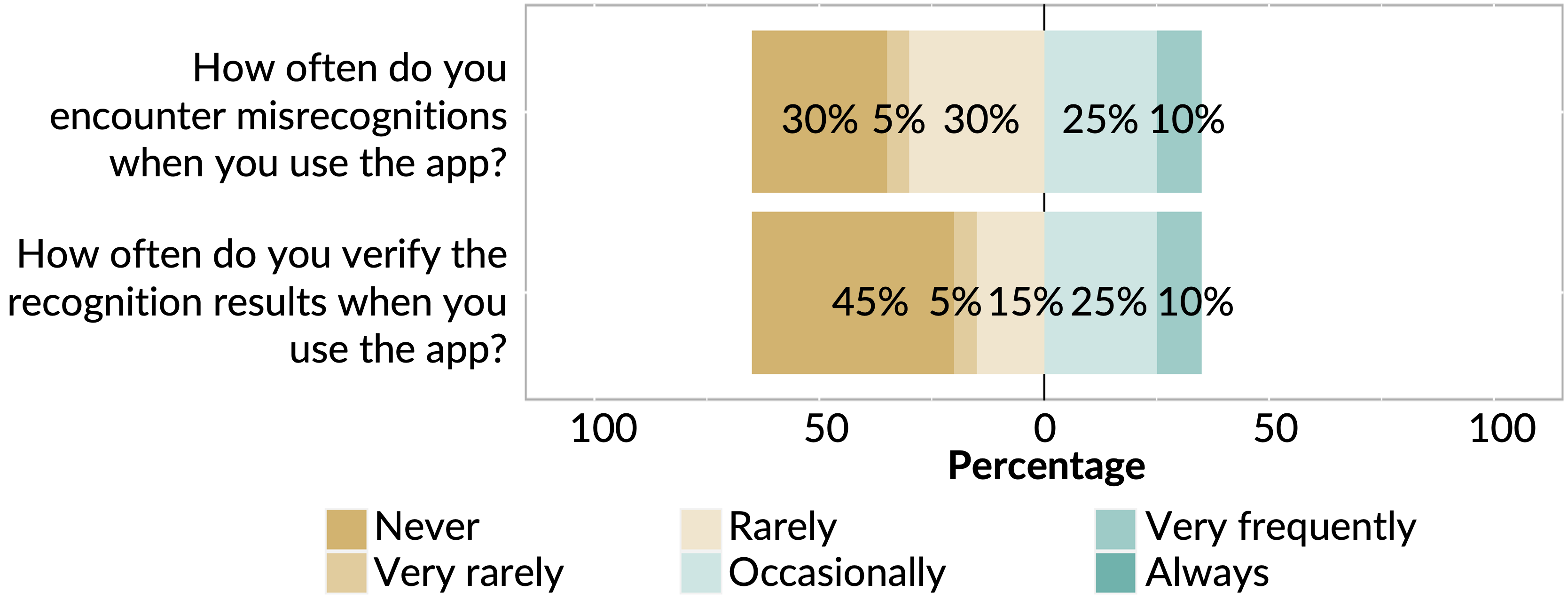}
    \caption{Participants' responses about frequency of encountered errors and verification of the output from the apps.}
    \label{fig:tor1-interview-apps-errors1}
    \Description{A grouped horizontal bar chart depicting responses to two questions about app usage habits. The questions are: 'How often do you encounter misrecognitions when you use the app?' and 'How often do you verify the recognition results when you use the app?'. The x-axis represents the percentage of respondents, ranging from 0\% to 100\%. The y-axis displays the frequency categories: 'Never', 'Very rarely', 'Rarely', 'Occasionally', 'Very frequently', 'Always'. The bars are color-coded to represent different response frequencies. The values displayed on the bars are as follows:

    'How often do you encounter misrecognitions when you use the app?':
    Never: 30\%
    Very rarely: 5\%
    Rarely: 30\%
    Occasionally: 25\%
    Very frequently: 10\%
    Always: 0\%
    'How often do you verify the recognition results when you use the app?':
    Never: 45\%
    Very rarely: 5\%
    Rarely: 15\%
    Occasionally: 25\%
    Very frequently: 10\%
    Always: 0\%}
\end{figure}

To gain insights into the experiences and preferences regarding camera-based assistive applications, we conducted a comprehensive inquiry into the apps they regularly utilize. Participants reported using a total of eight camera-based assistive apps, with inquiries aimed at elucidating their experiences with each. Across 20 participant-app pairs, the predominant choice was Seeing AI, as depicted in Figure~\ref{fig:tor1-interview-apps}. Additionally, participants employed other apps offering text and object recognition functionalities, including Google Lookout, KNFB Reader, Super Lidar, Supersense, and Voice Dream Scanner. Aira and Be My Eyes were also utilized for obtaining remote sighted assistance. Participants varied in the frequency of app usage, with some employing them several times a day ($N=5$), several times a week ($N=7$), several times a month ($N=5$), or once a month ($N=3$). When asked about the frequency of encountering misrecognitions, responses varied, ranging from very frequently ($N=2$) and occasionally ($N=5$) to rarely ($N=6$), very rarely ($N=1$), and never ($N=6$), as shown in Figure~\ref{fig:tor1-interview-apps-errors1}. However, it's noteworthy that participants might not have perceived all errors. Thus, the reported frequency of errors could be lower than the actual frequency. 

We also inquired about participants' strategies for capturing ``good'' photos when using each camera-based assistive app. To capture quality photos, participants employed strategies such as adjusting the distance and orientation of the camera ($N=9$ and $N=7$, respectively) and centering objects in the camera frame ($N=7$). This reflects the perceived challenge of image framing mentioned earlier. Additionally, participants utilized computer-generated feedback for blind photography ($N=8$), such as the audio tone system described by P12, a user of Voice Dream Scanner, who stated, \textit{``It has this system where the louder and steadier the audio tone is, the better you are. There's a certain tone. You've got the perfect picture and you snap it.''} P1 highlighted comparable feedback from Seeing AI for taking a photo of a person, stating, \textit{``I listen to the prompts. It'll tell me if the face is at the bottom left or top right. Or face is at center. When I hear that. That's when I push the button.''}

We delved deeper into how participants addressed potential recognition errors they may have experienced with these apps. We queried participants about the frequency with which they validated predictions from the apps (Figure~\ref{fig:tor1-interview-apps-errors1}). In the majority of cases, participants reported never verifying outputs while using the apps ($N=9$). Many of them expressed trust in the app's outputs without validation ($N=7$), exemplified by statements such as \textit{``if it says it's a \$5 bill, I believe it''} (P2, Seeing AI), \textit{``I assume it's correct when it reads it to me''} (P6, Seeing AI), and \textit{``(I rarely verify the recognition results) because it's pretty accurate.''} (P12, Google Lookout). This response aligns with findings from prior studies indicating that blind users tend to trust computer-vision systems~\cite{macleod2017understanding}. Some participants refrained from validating outputs because they found errors easy to detect ($N=6$). Particularly with text recognition apps, they could identify errors if the outputs did not make sense. For instance, P11, who never verified outputs from Seeing AI and Voice Dream Scanner, stated \textit{``If it tells me a certain thing, I'll know that it actually meant certain numbers. The errors that are sometimes made, they kind of have patterns if you know what it is.''} This response aligns with the findings of a study by Guerreiro \etal~\cite{guerreiro2018context}, which suggests that errors are often acceptable when users understand the imperfections of the technology. When recognizing objects, participants compared app outputs with their expectations based on object textures, shapes, and weights. For instance, P6, who never validated outputs from Seeing AI, mentioned \textit{``[...] I could say sometimes it does get the canned soup name wrong, but I guess I don't consider it wrong enough to call it wrong.''} Some participants verified outputs occasionally ($N=5$), rarely ($N=3$), or very rarely ($N=1$). The most common reason for verifying results was uncertainty with a single output, prompting the need for multiple trials to make a decision ($N=8$). For example, P3 explained, \textit{``if I'm consistently not getting a result with Seeing AI, then I'll see if KNFB Reader will give me results.''}

\begin{figure}[t]
    \centering
    \includegraphics[width=1\columnwidth]{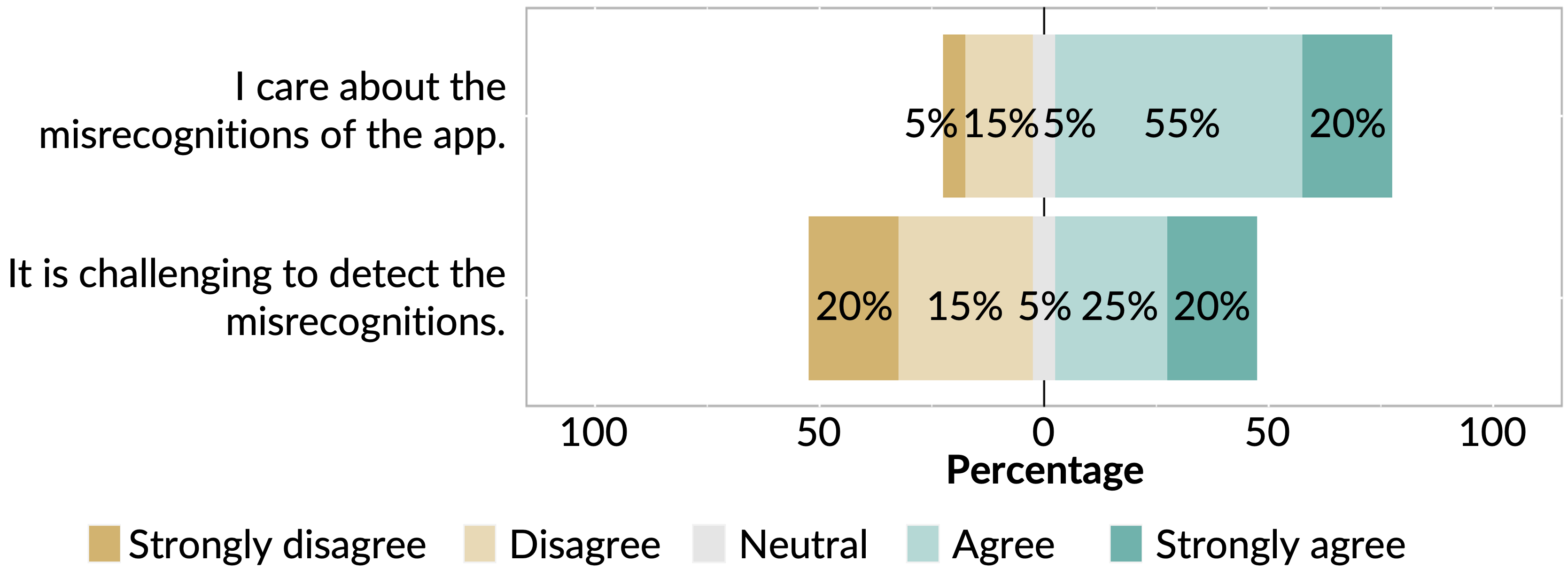}
    \caption{Participants' responses about handling errors.}
    \label{fig:tor1-interview-apps-errors2}
    \Description{A grouped horizontal bar chart displaying responses to two questions. The questions are: 'I care about the misrecognitions of the app.' and 'It is challenging to detect the misrecognitions.'. The x-axis represents the percentage of respondents, ranging from 0\% to 100\%. The y-axis displays the response options: 'Strongly disagree', 'Disagree', 'Neutral', 'Agree', 'Strongly agree'. The bars are color-coded to represent different response categories. The values displayed on the bars are as follows:

    'I care about the misrecognitions of the app.':
    Strongly disagree: 5\%
    Disagree: 15\%
    Neutral: 5\%
    Agree: 55\%
    Strongly agree: 20\%
    'It is challenging to detect the misrecognitions.':
    Strongly disagree: 20\%
    Disagree: 15\%
    Neutral: 5\%
    Agree: 25\%
    Strongly agree: 35\%"}   
\end{figure}

\textbf{While the frequency of encountering errors varied among the participants, the majority expressed concern about the misrecognitions.} We delved into the impact of errors in camera-based assistive technology on users' experiences with it. In most cases, participants either agreed ($N=13$) or strongly agreed ($N=3$) that they cared about the misrecognitions from the apps, as depicted in Figure~\ref{fig:tor1-interview-apps-errors2}. Sometimes, however, they did not prioritize error correction because they could understand the outputs even with some errors. For instance, errors in text recognition did not significantly alter the meaning of the texts, or the apps were not utilized for sensitive or critical tasks. P8 (Seeing AI) expressed this sentiment, stating, \textit{``It's not the most important thing, because I'm not using it for something critical.''} When asked if there were situations where they cared more about errors, participants often cited text recognition scenarios involving important content such as bills, currency, expiration dates, or other crucial numbers ($N=11$). For instance, P1, a Be My Eyes user, explained,\textit{ ``if they don't see the expiration date properly on something and it's expired, you know, I could get sick.''} Other critical situations included reading directions for tasks ($N=5$) and reviewing important documents ($N=5$). P9 (Voice Dream Scanner) provided examples of such documents, stating, `\textit{`probably when it's something that is connected to legal documents, financial statements, legal financial statements.''} Responses regarding the difficulty of identifying misrecognitions varied. \textbf{In half of all cases ($N=10$), participants disagreed or strongly disagreed that identifying errors was challenging when they could easily detect them using contextual cues such as surrounding text or object textures.} For example, P1 (Be My Eyes) remarked, \textit{``if they're wrong, I know they're wrong. So it's not really a challenge to identify that it's a misrecognition for me.''} P12 (Seeing AI) similarly commented, \textit{``I can catch the errors as they come up because often, it's not wrong enough for me to not be able to figure out what it says.''} Conversely, in other cases, participants ($N=9$) found errors less distinguishable and challenging to identify. P8 (Seeing AI), acknowledging the possibility of missing errors, expressed, \textit{``If it's wrong, I wouldn't know. [...] I don't even know whether it's wrong or true.''} Additionally, P9 recounted instances where sighted individuals detected errors from Voice Dream Scanner that she had missed, stating, \textit{``There have been occasions when I didn't detect anything and a sighted person may have indicated there was something that I just did not get.''} When asked how they identified errors, the majority ($N=10$) of participants relied on contextual cues. For example, P1, using Seeing AI for text recognition, mentioned, \textit{``If the information reading isn't very clear, if I can tell that it's only reading a part of something then I have to readjust it.''} Similarly, P6, identifying objects with Seeing AI, explained, \textit{``if I get a soup, and it's not pronouncing the type of soup, that type of thing.''} This behavior contrasts with the strategies of blind users in handling errors in navigation systems, where the majority sought sighted assistance when they encountered errors~\cite{guerreiro2018context}. In other cases, participants sought clarification from sighted individuals ($N=5$) or verified app outputs through multiple trials ($N=5$).

\section{Error Identification Results}
We conducted a comprehensive evaluation of participants' experience with identifying errors within the context of object recognition. Our analysis centered on discerning patterns in participants' error-handling behavior throughout the task. Additionally, we examined the influence of repeated object recognition efforts on error handling by comparing the two attempts. Furthermore, participants' feedback provided valuable insights into their attitudes toward errors encountered in object recognition.

\subsection{Identifying Object Recognition Errors}
\label{error_identificaiton_task_results_1}

Across the 30 trials (15 in the first and 15 in the second attempt), the average accuracy of object recognition stood at 0.76 ($SD=0.10$). Participants encountered 7.33 incorrect recognitions on average ($SD=2.99$), experiencing more false positives ($M=3.67$, $SD=2.46$) than false negatives ($M=0.83$, $SD=1.03$). 
When looking at whether participants could distinguish between correct and incorrect recognitions, we find that  on average, participants successfully identified 21.83 ($SD=2.82$) correct (true positives) and 3.17 ($SD=2.44$) incorrect (true negatives) recognition results. However, participants identified errors at a proportion of 0.49 on average ($SD=0.32$), indicating \textbf{that they could detect less than half of the errors}. This outcome is consistent with the over-reliance on image recognition results observed in a previous study by MacLeod \etal~\cite{macleod2017understanding}. The low rate of error identification can be attributed to the challenge of distinguishing objects within the same category that share similar shapes, textures, and weights (\eg, coca-cola and diet coke) when limited visual information is available.

\begin{figure}[t]
    \centering
    \includegraphics[width=\columnwidth]{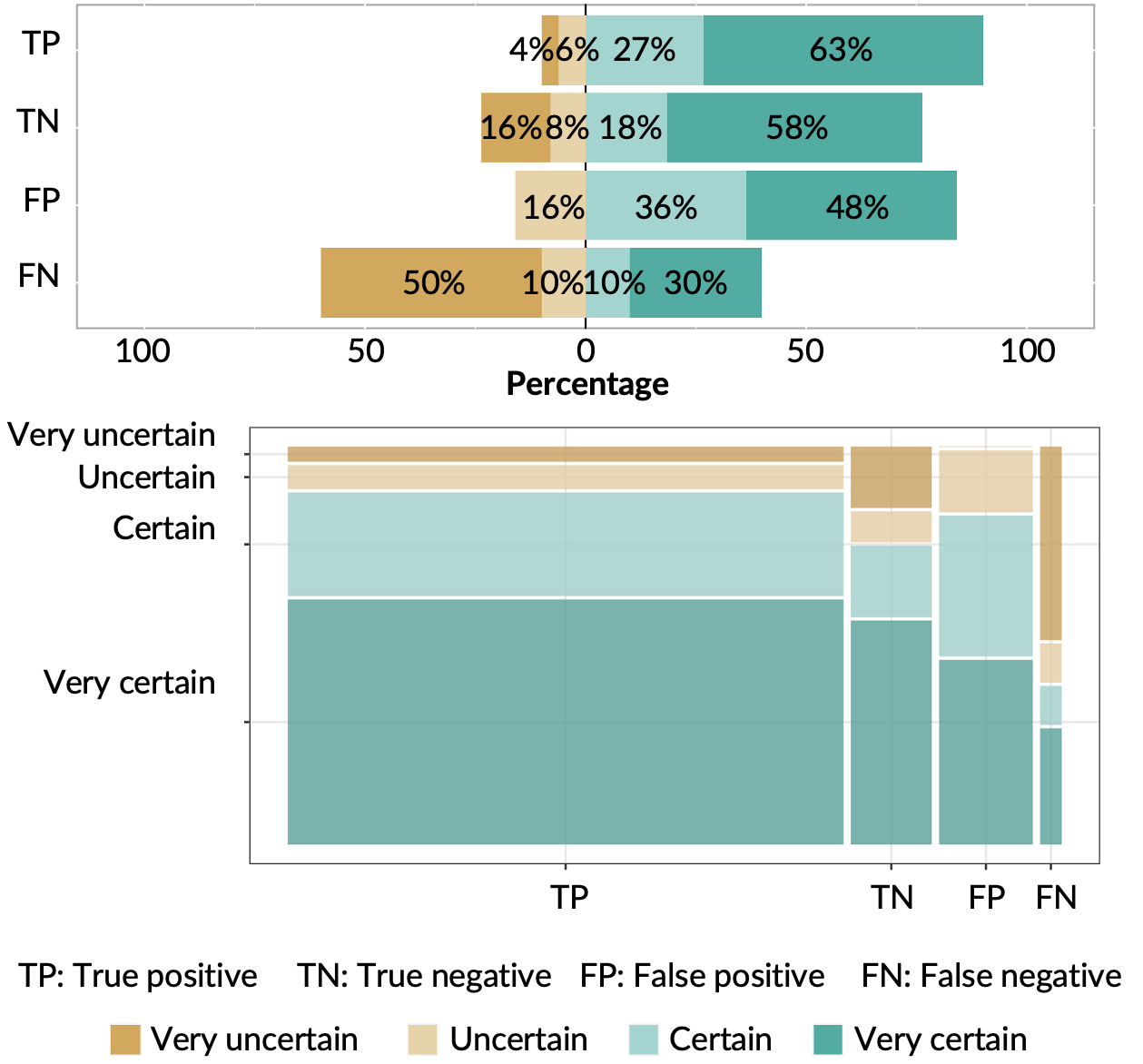}
    \caption{Likert chart (top) and mosaic plot (bottom) of certainty levels and trial categories. The size of the rectangles in the mosaic plot corresponds to the number of trials.}
    \label{fig:certainty_correctness}
    \Description{Left: A grouped horizontal bar chart displaying the categories of trials and level of certainty. The x-axis represents the percentage of respondents, ranging from 0\% to 100\%. The y-axis displays the categories of trials: 'true positive', 'true negative', 'false positive', and 'false negative'. The bars are color-coded to represent different levels of certainty. The values displayed on the bars are as follows:

    'true positive':
    Very uncertain: 4\%
    Uncertain: 6\%
    Certain: 27\%
    Very certain: 63\%

    'true negative':
    Very uncertain: 16\%
    Uncertain: 8\%
    Certain: 18\%
    Very certain: 58\%

    'false positive':
    Very uncertain: 0\%
    Uncertain: 16\%
    Certain: 36\%
    Very certain: 48\%

    'false negative':
    Very uncertain: 50\%
    Uncertain: 10\%
    Certain: 10\%
    Very certain: 30\%

    Right: mosaic plot showing the rectangles with different sizes. The level of certainty is on the y-axis. The categories of trials are on the x-axis.
    }
\end{figure}

\begin{table}
\small
\centering
\caption{Participants' strategies to overcome \textit{``Don't know''}.}
       \setlength\tabcolsep{1pt} 
       \begin{tabular}{p{1.6cm} p{5.7cm} c}
            \toprule              
            \textbf{Code}  & \textbf{Strategy} & \textbf{Cases} \\
            \hline
            \textbf{Object size}  & Adjust the camera distance for better framing & 11 \\ 
            \hline
            \multirow{3}{*}{\textbf{Background}}  
            & Rotate the object to show different sides & 119 \\
            & Move the object to another place & 23 \\
            & Hide the background objects with a paper & 1 \\
            \hline
            \multirow{4}{*}{\textbf{Viewpoint}}
            & Move the camera to display other sides of the object & 29 \\
            & Change the way of holding the object & 17 \\
            & Rotate the object to change perspective & 5 \\
            & Rotate the camera (portrait and landscape) & 1 \\
            \hline
            & No change & 10 \\
            \bottomrule
        \end{tabular}%
\label{tab:variations_dn}
\end{table}


When looking at participants' strategies for recovery from \textit{``Don't know''} predictions, we find that they cluster around varying object size, background, and viewpoint (shown in Table~\ref{tab:variations_dn}). This is exciting as none of the participants reported having machine learning expertise. Yet, these patterns underscore \textbf{participants' awareness of the potential impact of object's size, viewpoint and background on the performance of the object recognition model}, drawing from parallels to how humans recognize objects independent of size, viewpoint, location, and illumination~\cite{palmeri2004visual}.  On average, the object recognition app provided a \textit{``Don't know''} response in 8.2 trials ($SD=4.17$) out of 30 trials, totaling 216 cases;  a \textit{``Don't know''} response would often be followed by subsequent a \textit{``Don't know''} responses with an average of 2.01 ($SD=0.89$) occurrences. As detailed in Table~\ref{tab:variations_dn}, when participants encountered \textit{``Don't know,''} the most prevalent (116 cases) approach to circumvent it was rotating the object to display its other side, thereby varying the viewpoint in the image, a strategy also prevalent among sighted non-experts in prior work~\cite{hong2020crowdsourcing}. The second most common approach (29 cases) also involved adjusting the viewpoint, with participants moving the camera instead of the object. Additionally, participants occasionally (23 cases) altered the background of the image by relocating the object to different positions. 

We examine participants certainty around the error identification task by looking at their responses for each trial where they indicate their confidence in their judgment of the model prediction. Overall, participants expressed varying levels of certainty, reporting being ``very certain,'' ``certain,'' ``uncertain,'' and ``very uncertain'' across 17.67 ($SD=5.71$), 7.83 ($SD=5.77$), 2.25 ($SD=2.83$), and 1.75 ($SD=2.01$) trials, respectively. As shown in Figure~\ref{fig:certainty_correctness}, we find that participants reported being either certain or very certain in 90\% of true positive trials; trials where the object recognition is correct and the participant perceive it as such. This seems promising. Yet, they also reported being either certain or very certain in
84\% of false positive trials; trials where the object recognition is incorrect but the participant perceive it as such. In contrast, participants reported being either certain or very certain in 76\% of true negative trials, and 40\% of false negative trials. This trend underscores \textbf{a tendency for heightened certainty when participants perceived recognition outcomes as correct}. 
Overall, these findings indicate a prevalent inclination among participants to place trust in the predictions from the object recognizer.

Through analysis of trial completion time, we observe that \textbf{participants tend to make quicker decisions regarding the correctness of predictions when they were very certain} ($M=3.91s, SD=2.70$) compared to when they were just certain ($M=8.48s, SD=5.28$), uncertain ($M=7.87s, SD=2.71$), or very uncertain ($M=7.69s, SD=8.30$). A small correlation was observed between the level of certainty and the trial completion time, as indicated by the Pearson Correlation Coefficient ($r=0.27$).

\begin{figure}[b]
    \centering
    \includegraphics[width=\columnwidth]{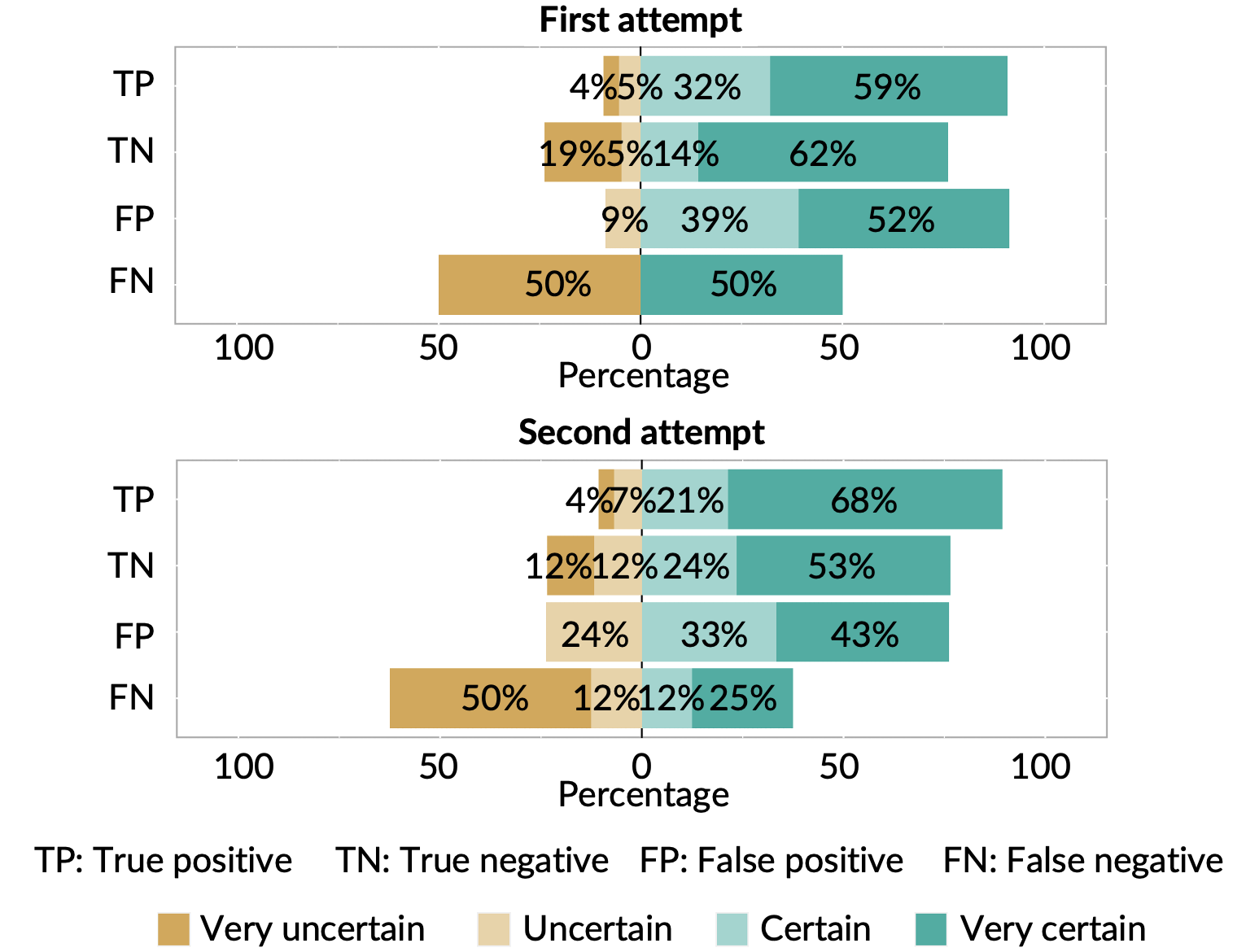}
    \caption{Percentages of the certainty levels across the categories of the trials in the first and second attempts.}
    \label{fig:certainty_correctness_attempts}
    \Description{A grouped horizontal bar chart comparing the percentages of true positive (TP), true negative (TN), false positive (FP), and false negative (FN) cases in four categories of trials between the first attempt and second attempt. The percentages are represented in the following order: very uncertain, uncertain, certain, very certain. For the first attempt, TP percentages are 4\% 5\%, 32\%, and 59\%, TN percentages are 19\%, 5\%, 14\%, and 62\%, FP percentages are 0\%, 9\%, 39\%, and 52\%, and FN percentage is 50\%, 0\%, 0\%, and 50\%. For the second attempt, TP percentages are 4\%, 7\%, 21\%, and 68\%, TN percentages are 12\%, 12\%, 24\%, and 53\%. FP percentages are 0\%, 24\%, 32\%, and 43\%, and FN percentages are 50\%, 12\%, 12\%, and 25\%. The x-axis represents the percentage scale from 0\% to 100\%. Each bar is color-coded to represent the respective category.}
\end{figure}

\subsection{Identifying Errors a Second Time}
In a real-world scenario, participants tend to interact with a recognition application and similar objects over a long period and often learn to anticipate failures. In Section~\ref{error_identificaiton_task_results_1}, we present aggregated observations from both attempts. To understand even at a small scale the effect of repeated use of the object recognition application on handling incorrect recognitions, in this section we compared the two attempts in the error identification task. Overall, we find that \textbf{the proportion of errors identified by the participants was not significantly\footnote{We did not observe a statistically significant difference in the results of repeated measures Analysis of Variance (ANOVA) with Aligned Rank Transform (ART) regarding the number of errors and the proportion of errors ($p>.05$).} different across the two} with it being at 0.51 on average  ($SD=0.40$) for the first and 0.46 ($SD=0.36$) for the second attempt.  Regarding the level of certainty in the correctness of the recognitions, in the second attempt, \textbf{participants were certain or very certain for a smaller proportion of trials} across all four categories compared to the first attempt,  as shown in Figure~\ref{fig:certainty_correctness_attempts}. One of the reasons for this difference was inconsistent recognition results with the same object across the first and second attempts, supported by P9's response: \textit{``the second time around, they gave me different information. So then I became uncertain about trusting what it was telling me.''} 

Furthermore, in the second attempt, \textbf{the trial completion time significantly\footnote{The results of the repeated measures ANOVA with ART exhibited a statistically significant difference ($F_{1, 9}=9.67, p=.013, \eta^2=0.52$).} decreased} to 4.22 seconds ($SD=2.33$), compared to the first, where we recorded a longer duration of 6.75 seconds ($SD=3.15$).  This discrepancy suggests a meaningful variation between the attempts. Possible explanations for this observed difference could be attributed to participants' increased familiarity with the task procedure in the second attempt, as well as quicker decision-making based on prior experience with the task in the first attempt.

\subsection{Subjective Feedback}
While participants missed around half of the errors, they generally perceived identifying errors as not challenging, confirming the finding from a prior study that BLV users have mixed feelings with both confidence and concerns regarding identifying errors in private object detection~\cite{zhang2024designing}. When asked about the difficulty, the majority disagreed ($N=5$), with some strongly disagreeing ($N=3$). For instance, P8, who has low vision, was able to discern correct and incorrect predictions based on their vision and the textures of the object. Other participants identified errors by comparing predictions across multiple trials. For example, P10 explained, \textit{``I didn't recognize a mistake until the second similar object appeared. So like the two cans of the Lacroix apricot and Lacroix mango, one of them was incorrect because it was telling me apricot both times.''} Errors were sometimes evident to participants because predicted and true objects had distinct textures, shapes, or weights, as noted by P12: \textit{``[...] for example, the diced tomatoes versus the chicken broth, chicken broth is more liquid. It was easy to identify that it was wrong.''} On the other hand, three participants strongly agreed that identifying errors was challenging. Among them, two mentioned that the recognition results were inconsistent with an object, making it difficult to determine their correctness. P9 said \textit{``Two things that seem similar, but the first time they said they were the same, and then the next time putting them back, they said something different on one of them. So now I’m not sure. So I strongly agree, it was difficult for me to tell us it was in error.''} Another participant mentioned that it was challenging to remember all objects explained at the beginning of the study, which complicated the decision-making regarding the correctness of recognition results.

\section{Discussion}
Our user study, exploratory in nature, shows both promising results and future research directions for supporting blind users' interactions with error-prone AI-infused technologies. In this section we discuss lessons learned and limitations  that may affect the generalizability of our findings.

\subsection{Implications}

\textbf{Enable users to leverage their expertise in reviewing errors independently.}
The findings from the interview have shed light on an interesting trend: most participants expressed a preference for evaluating the quality of their photographs without the assistance of sighted individuals or remote sighted aid services, such as Be My Eyes or Aira, when using camera-based assistive technologies. This preference seems to stem from a fundamental aspect of the utilization of AI-based systems -- namely, the desire to carry out visual tasks independently when sighted assistance is unavailable. It further shows the preference of blind and low-vision users to utilize their expertise in assistive technology, such as integrating recognition results from multiple AI apps to identify errors. This personalized approach to using assistive technology was highlighted in a previous study~\cite{herskovitz2023hacking}. This observation underscores a crucial need within the blind community: the ability for individuals to autonomously assess the quality of their photos, taking into account factors such as framing, background clutter, and blurriness.

Addressing this challenge will likely require innovative approaches, particularly in the realm of computer vision. Developing techniques that can accurately quantify the quality factors of photographs without relying on visual cues accessible only to sighted individuals holds great promise in this regard. Such techniques could potentially leverage advanced algorithms and machine learning models to analyze various aspects of a photograph, from composition to sharpness, and provide meaningful feedback to blind and low-vision users. While some initial strides have been made in this area, such as image descriptors for blind users to assess photos for training personalized object recognition systems~\cite{hong2022blind} and real-time feedback for blind users to capture high-quality photos~\cite{mandal2023helping, ahmetovic2020recog}, there remains a need for further investigation. Specifically, it is essential to evaluate the effectiveness of these descriptors in identifying errors and providing actionable insights to users. 

\textbf{Incorporate the context and recognition system type in designing intuitive user interfaces.}
In our interviews, participants delineated diverse approaches to pinpointing errors in both object and text recognition. When utilizing text recognition, they relied heavily on contextual cues, as errors often manifested as deviations from the surrounding text's logical flow. In contrast, with object recognition, participants leveraged intrinsic object properties such as weight and texture to gauge recognition accuracy. Additionally, certain applications with vision language models like Be My AI~\cite{BeMyAI} furnish detailed image descriptions, enriching user experience. However, this potentially introduces more complex challenges in error detection due to the longer and more descriptive texts~\cite{amin2023modeling}, compared to the simple object labels in URCam. While our study primarily delved into object recognition, participant feedback underscores the pivotal role of recognition system type and contextual understanding in crafting user interfaces for error detection in camera-based assistive technologies. 

Consequently, our findings offer valuable insights for designing intuitive interfaces tailored to object recognition error identification with images. For instance, our findings suggest that providing descriptive information about different facets of an object in an image could mitigate error occurrences, as evidenced by participants frequently resorting to rotating objects to avoid \textit{"Don't know"} response from the URCam app during the error identification tasks. On the other hand, real-time camera-based assistive technologies may present unique challenges. We expect that blind users would employ different strategies for avoiding and validating errors since they can observe the effects of their camera framing immediately. This immediate feedback loop could encourage adaptive behaviors, such as repositioning the camera or altering the angle of capture to ensure better recognition accuracy. Future research should explore these adaptive strategies in depth, examining how real-time feedback influences user interaction patterns and error mitigation techniques.


\textbf{Enable users to understand the performance of the object recognizer.}
While participants missed approximately half of the errors, their collective perception of error identification as non-challenging was notable. When probed about the difficulty level, the majority of participants disagreed, with some expressing strong disagreement. This observation underscores the nuanced difficulty inherent in comprehending both the overall performance metrics of the object recognizer (\ie, error rate) and pinpointing individual errors, a challenge compounded for blind and low-vision users. This corroborates findings from prior studies that showed the blind and low-vision users' tendency to exhibit an overtrust on the output AI-based assistive technologies such as image recognition systems~\cite{macleod2017understanding} and automatic speech recognition~\cite{hong2020reviewing}.

While studies within the domain of Explainable AI have demonstrated the potential efficacy of elucidating the certainty and rationale behind machine learning model outputs in enhancing performance understanding and usability~\cite{shin2021effects, vorm2018assessing}, many of these studies rely on visual information such as heatmap~\cite{jin2023guidelines} and plots~\cite{pradhan2023explainable} inaccessible to blind users or have not been assessed with blind individuals. Consequently, to facilitate error identification effectively, forthcoming research endeavors must prioritize the development of methodologies that enable blind and low-vision users to assess the performance of object recognition systems.

\subsection{Limitations}

\textbf{Variability between confined study conditions and real-world experience.}
A notable limitation of our study lies in the potential disparity between error identification under confined conditions and real-world usage scenarios. In the user study setting, participants were confined to a specific study setup in their home environment with limited variables such as lighting, background, and framing. Typically, they would place the study materials on a table and sit nearby. In a way, they were restricted in their ability to move around freely to find optimal positions for capturing photos, which could influence the quality of the image and subsequently impact error identification. Furthermore, a somewhat `staged' indoor setting may not fully replicate the diverse conditions encountered in real-world scenarios, such as varying lighting conditions, backgrounds, and the presence of outdoor elements. Participants' level of familiarity and experience with the application may also differ between a one-off and real-world usage contexts. While participants received guidance and instructions during the user study, their experience in using the application in real-world settings may vary, potentially affecting their proficiency in error identification. Last, the number and types of objects encountered in real-world scenarios may differ from the stimuli in the study. Real-world scenarios often involve a wider variety of objects and contexts, presenting unique challenges for error identification.

\textbf{Single-session limitation and potential longitudinal variability.}
An inherent limitation of our study is that the error identification task was conducted within a single session at participants' homes. While this approach allowed us to gather valuable data in a naturalistic setting, it may not fully capture the evolution of participants' error identification abilities over time. Indeed, our observations revealed differences between participants' performance in the first and second attempts of the error identification task. This discrepancy suggests that participants' understanding of the object recognizer's performance, the characteristics of objects, and optimal photo-taking techniques may have improved with repeated exposure and experience. Consequently, a longitudinal study spanning multiple sessions could provide deeper insights into how participants' error identification abilities evolve over time.

Therefore, while our study provides valuable initial insights into error identification in a single-session context, future research employing longitudinal methodologies could offer a more comprehensive understanding of the development and refinement of error identification experience and expertise that blind users build while interacting with their camera-based assistive technologies.

\section{Conclusion}
We explored the experiences of blind and low-vision people regarding photo-taking, usage of camera-based assistive systems, and error identification within these systems. Through semi-structured interviews, we uncovered that participants predominantly utilize photo-taking for the purpose of utilizing camera-based assistive systems, rather than solely for capturing memories or sharing with others. Additionally, participants revealed their inclination towards independently reviewing photo quality and identifying errors, despite acknowledging the challenging nature of these tasks for approximately half of the participants. Furthermore, our empirical investigation through error identification tasks provided valuable insights into the challenges associated with identifying object recognition errors. The results indicated that participants successfully identified only around 50\% of the errors, predominantly employing viewpoint, background, and object size alterations within images to mitigate errors. Additionally, we observed that the certainty regarding recognition correctness could be adversely affected by inconsistent recognition outcomes in subsequent interactions. These findings significantly contribute to our understanding and quantification of the challenges in identifying object recognition errors within assistive technologies.


\begin{acks}
We thank Kyungjun Lee, Ebrima Jarjue, and Ernest Essuah Mensah, who were students at the University of Maryland at the time of the data collection and contributed to the remote study protocol. Jonggi Hong initiated this work at the University of Maryland, College Park. This material is based upon work supported by the National Science Foundation under Grant No. 1816380. Hernisa Kacorri was additionally supported by the National Institute on Disability, Independent Living, and Rehabilitation Research (NIDILRR), ACL, HHS under Grant No. 90REGE0008 and 90REGE0024.
\end{acks}

\bibliographystyle{ACM-Reference-Format}
\bibliography{references}


\begin{thebibliography}{92}


\ifx \showCODEN    \undefined \def \showCODEN     #1{\unskip}     \fi
\ifx \showDOI      \undefined \def \showDOI       #1{#1}\fi
\ifx \showISBNx    \undefined \def \showISBNx     #1{\unskip}     \fi
\ifx \showISBNxiii \undefined \def \showISBNxiii  #1{\unskip}     \fi
\ifx \showISSN     \undefined \def \showISSN      #1{\unskip}     \fi
\ifx \showLCCN     \undefined \def \showLCCN      #1{\unskip}     \fi
\ifx \shownote     \undefined \def \shownote      #1{#1}          \fi
\ifx \showarticletitle \undefined \def \showarticletitle #1{#1}   \fi
\ifx \showURL      \undefined \def \showURL       {\relax}        \fi
\providecommand\bibfield[2]{#2}
\providecommand\bibinfo[2]{#2}
\providecommand\natexlab[1]{#1}
\providecommand\showeprint[2][]{arXiv:#2}

\bibitem[Abdolrahmani et~al\mbox{.}(2017)]%
        {abdolrahmani2017embracing}
\bibfield{author}{\bibinfo{person}{Ali Abdolrahmani}, \bibinfo{person}{William
  Easley}, \bibinfo{person}{Michele Williams}, \bibinfo{person}{Stacy Branham},
  {and} \bibinfo{person}{Amy Hurst}.} \bibinfo{year}{2017}\natexlab{}.
\newblock \showarticletitle{Embracing errors: Examining how context of use
  impacts blind individuals' acceptance of navigation aid errors}. In
  \bibinfo{booktitle}{\emph{Proceedings of the 2017 CHI Conference on Human
  Factors in Computing Systems}}. \bibinfo{pages}{4158--4169}.
\newblock


\bibitem[Afzal et~al\mbox{.}(2020)]%
        {afzal2020study}
\bibfield{author}{\bibinfo{person}{Afsoon Afzal}, \bibinfo{person}{Deborah~S
  Katz}, \bibinfo{person}{Claire~Le Goues}, {and}
  \bibinfo{person}{Christopher~S Timperley}.} \bibinfo{year}{2020}\natexlab{}.
\newblock \showarticletitle{A study on the challenges of using robotics
  simulators for testing}.
\newblock \bibinfo{journal}{\emph{arXiv preprint arXiv:2004.07368}}
  (\bibinfo{year}{2020}).
\newblock


\bibitem[Afzal et~al\mbox{.}(2021)]%
        {afzal2021simulation}
\bibfield{author}{\bibinfo{person}{Afsoon Afzal}, \bibinfo{person}{Deborah~S
  Katz}, \bibinfo{person}{Claire Le~Goues}, {and}
  \bibinfo{person}{Christopher~S Timperley}.} \bibinfo{year}{2021}\natexlab{}.
\newblock \showarticletitle{Simulation for robotics test automation: Developer
  perspectives}. In \bibinfo{booktitle}{\emph{2021 14th IEEE conference on
  software testing, verification and validation (ICST)}}. IEEE,
  \bibinfo{pages}{263--274}.
\newblock


\bibitem[Ahmetovic et~al\mbox{.}(2020)]%
        {ahmetovic2020recog}
\bibfield{author}{\bibinfo{person}{Dragan Ahmetovic}, \bibinfo{person}{Daisuke
  Sato}, \bibinfo{person}{Uran Oh}, \bibinfo{person}{Tatsuya Ishihara},
  \bibinfo{person}{Kris Kitani}, {and} \bibinfo{person}{Chieko Asakawa}.}
  \bibinfo{year}{2020}\natexlab{}.
\newblock \showarticletitle{Recog: Supporting blind people in recognizing
  personal objects}. In \bibinfo{booktitle}{\emph{Proceedings of the 2020 CHI
  Conference on Human Factors in Computing Systems}}. \bibinfo{pages}{1--12}.
\newblock


\bibitem[AI(2024)]%
        {BeMyAI}
\bibfield{author}{\bibinfo{person}{Be~My AI}.} \bibinfo{year}{2024}\natexlab{}.
\newblock \bibinfo{booktitle}{\emph{Introducing: Be My AI}}.
\newblock
\urldef\tempurl%
\url{https://www.bemyeyes.com/blog/introducing-be-my-ai}
\showURL{%
\tempurl}


\bibitem[Aira(2024)]%
        {Aira}
\bibfield{author}{\bibinfo{person}{Aira}.} \bibinfo{year}{2024}\natexlab{}.
\newblock \bibinfo{booktitle}{\emph{Your Life, Your Schedule, Right Now.}}
\newblock
\urldef\tempurl%
\url{https://aira.io}
\showURL{%
\tempurl}


\bibitem[Ajenaghughrure et~al\mbox{.}(2020)]%
        {ajenaghughrure2020risk}
\bibfield{author}{\bibinfo{person}{Ighoyota~Ben Ajenaghughrure},
  \bibinfo{person}{Sonia~Claudia da Costa~Sousa}, {and} \bibinfo{person}{David
  Lamas}.} \bibinfo{year}{2020}\natexlab{}.
\newblock \showarticletitle{Risk and Trust in artificial intelligence
  technologies: A case study of Autonomous Vehicles}. In
  \bibinfo{booktitle}{\emph{2020 13th International Conference on Human System
  Interaction (HSI)}}. IEEE, \bibinfo{pages}{118--123}.
\newblock


\bibitem[Akter et~al\mbox{.}(2020)]%
        {akter2020uncomfortable}
\bibfield{author}{\bibinfo{person}{Taslima Akter}, \bibinfo{person}{Bryan
  Dosono}, \bibinfo{person}{Tousif Ahmed}, \bibinfo{person}{Apu Kapadia}, {and}
  \bibinfo{person}{Bryan Semaan}.} \bibinfo{year}{2020}\natexlab{}.
\newblock \showarticletitle{" I am uncomfortable sharing what I can't see":
  Privacy Concerns of the Visually Impaired with Camera Based Assistive
  Applications}. In \bibinfo{booktitle}{\emph{29th USENIX Security Symposium
  (USENIX Security 20)}}. \bibinfo{pages}{1929--1948}.
\newblock


\bibitem[Alcorn et~al\mbox{.}(2019)]%
        {alcorn2019strike}
\bibfield{author}{\bibinfo{person}{Michael~A Alcorn}, \bibinfo{person}{Qi Li},
  \bibinfo{person}{Zhitao Gong}, \bibinfo{person}{Chengfei Wang},
  \bibinfo{person}{Long Mai}, \bibinfo{person}{Wei-Shinn Ku}, {and}
  \bibinfo{person}{Anh Nguyen}.} \bibinfo{year}{2019}\natexlab{}.
\newblock \showarticletitle{Strike (with) a pose: Neural networks are easily
  fooled by strange poses of familiar objects}. In
  \bibinfo{booktitle}{\emph{Proceedings of the IEEE/CVF Conference on Computer
  Vision and Pattern Recognition}}. \bibinfo{pages}{4845--4854}.
\newblock


\bibitem[Alharbi et~al\mbox{.}(2022)]%
        {alharbi2022understanding}
\bibfield{author}{\bibinfo{person}{Rahaf Alharbi}, \bibinfo{person}{Robin~N
  Brewer}, {and} \bibinfo{person}{Sarita Schoenebeck}.}
  \bibinfo{year}{2022}\natexlab{}.
\newblock \showarticletitle{Understanding emerging obfuscation technologies in
  visual description services for blind and low vision people}.
\newblock \bibinfo{journal}{\emph{Proceedings of the ACM on Human-Computer
  Interaction}} \bibinfo{volume}{6}, \bibinfo{number}{CSCW2}
  (\bibinfo{year}{2022}), \bibinfo{pages}{1--33}.
\newblock


\bibitem[Amin et~al\mbox{.}(2023)]%
        {amin2023modeling}
\bibfield{author}{\bibinfo{person}{Akhter~Al Amin}, \bibinfo{person}{Saad
  Hassan}, \bibinfo{person}{Matt Huenerfauth}, {and}
  \bibinfo{person}{Cecilia~Ovesdotter Alm}.} \bibinfo{year}{2023}\natexlab{}.
\newblock \showarticletitle{Modeling Word Importance in Conversational
  Transcripts: Toward improved live captioning for Deaf and hard of hearing
  viewers}. In \bibinfo{booktitle}{\emph{Proceedings of the 20th International
  Web for All Conference}}. \bibinfo{pages}{79--83}.
\newblock


\bibitem[Andreopoulos and Tsotsos(2013)]%
        {andreopoulos201350}
\bibfield{author}{\bibinfo{person}{Alexander Andreopoulos} {and}
  \bibinfo{person}{John~K Tsotsos}.} \bibinfo{year}{2013}\natexlab{}.
\newblock \showarticletitle{50 years of object recognition: Directions
  forward}.
\newblock \bibinfo{journal}{\emph{Computer vision and image understanding}}
  \bibinfo{volume}{117}, \bibinfo{number}{8} (\bibinfo{year}{2013}),
  \bibinfo{pages}{827--891}.
\newblock


\bibitem[Athavale et~al\mbox{.}(2020)]%
        {athavale2020ai}
\bibfield{author}{\bibinfo{person}{Jyotika Athavale}, \bibinfo{person}{Andrea
  Baldovin}, \bibinfo{person}{Ralf Graefe}, \bibinfo{person}{Michael
  Paulitsch}, {and} \bibinfo{person}{Rafael Rosales}.}
  \bibinfo{year}{2020}\natexlab{}.
\newblock \showarticletitle{AI and reliability trends in safety-critical
  autonomous systems on ground and air}. In \bibinfo{booktitle}{\emph{2020 50th
  Annual IEEE/IFIP International Conference on Dependable Systems and Networks
  Workshops (DSN-W)}}. IEEE, \bibinfo{pages}{74--77}.
\newblock


\bibitem[Bafghi and Gurari(2023)]%
        {bafghi2023new}
\bibfield{author}{\bibinfo{person}{Reza~Akbarian Bafghi} {and}
  \bibinfo{person}{Danna Gurari}.} \bibinfo{year}{2023}\natexlab{}.
\newblock \showarticletitle{A new dataset based on images taken by blind people
  for testing the robustness of image classification models trained for
  imagenet categories}. In \bibinfo{booktitle}{\emph{Proceedings of the
  IEEE/CVF Conference on Computer Vision and Pattern Recognition}}.
  \bibinfo{pages}{16261--16270}.
\newblock


\bibitem[BeMyEyes(2024)]%
        {BeMyEyes}
\bibfield{author}{\bibinfo{person}{BeMyEyes}.} \bibinfo{year}{2024}\natexlab{}.
\newblock \bibinfo{booktitle}{\emph{Lend you eyes to the blind}}.
\newblock
\urldef\tempurl%
\url{http://www.bemyeyes.org/}
\showURL{%
\tempurl}


\bibitem[Berke et~al\mbox{.}(2017)]%
        {berke2017deaf}
\bibfield{author}{\bibinfo{person}{Larwan Berke}, \bibinfo{person}{Christopher
  Caulfield}, {and} \bibinfo{person}{Matt Huenerfauth}.}
  \bibinfo{year}{2017}\natexlab{}.
\newblock \showarticletitle{Deaf and hard-of-hearing perspectives on imperfect
  automatic speech recognition for captioning one-on-one meetings}. In
  \bibinfo{booktitle}{\emph{Proceedings of the 19th International ACM SIGACCESS
  Conference on Computers and Accessibility}}. \bibinfo{pages}{155--164}.
\newblock


\bibitem[Braun and Clarke(2006)]%
        {braun2006using}
\bibfield{author}{\bibinfo{person}{Virginia Braun} {and}
  \bibinfo{person}{Victoria Clarke}.} \bibinfo{year}{2006}\natexlab{}.
\newblock \showarticletitle{Using thematic analysis in psychology}.
\newblock \bibinfo{journal}{\emph{Qualitative research in psychology}}
  \bibinfo{volume}{3}, \bibinfo{number}{2} (\bibinfo{year}{2006}),
  \bibinfo{pages}{77--101}.
\newblock


\bibitem[Brewer and Kameswaran(2018)]%
        {brewer2018understanding}
\bibfield{author}{\bibinfo{person}{Robin~N Brewer} {and}
  \bibinfo{person}{Vaishnav Kameswaran}.} \bibinfo{year}{2018}\natexlab{}.
\newblock \showarticletitle{Understanding the power of control in autonomous
  vehicles for people with vision impairment}. In
  \bibinfo{booktitle}{\emph{Proceedings of the 20th International ACM SIGACCESS
  Conference on Computers and Accessibility}}. \bibinfo{pages}{185--197}.
\newblock


\bibitem[Brinkley et~al\mbox{.}(2017)]%
        {brinkley2017opinions}
\bibfield{author}{\bibinfo{person}{Julian Brinkley}, \bibinfo{person}{Brianna
  Posadas}, \bibinfo{person}{Julia Woodward}, {and} \bibinfo{person}{Juan~E
  Gilbert}.} \bibinfo{year}{2017}\natexlab{}.
\newblock \showarticletitle{Opinions and preferences of blind and low vision
  consumers regarding self-driving vehicles: Results of focus group
  discussions}. In \bibinfo{booktitle}{\emph{Proceedings of the 19th
  International ACM SIGACCESS Conference on Computers and Accessibility}}.
  \bibinfo{pages}{290--299}.
\newblock


\bibitem[Brown(2010)]%
        {brown2010likert}
\bibfield{author}{\bibinfo{person}{Sorrel Brown}.}
  \bibinfo{year}{2010}\natexlab{}.
\newblock \showarticletitle{Likert scale examples for surveys}.
\newblock \bibinfo{journal}{\emph{ANR Program evaluation, Iowa State
  University, USA}} (\bibinfo{year}{2010}).
\newblock


\bibitem[Cao et~al\mbox{.}(2022)]%
        {cao2022s}
\bibfield{author}{\bibinfo{person}{Yang~Trista Cao}, \bibinfo{person}{Kyle
  Seelman}, \bibinfo{person}{Kyungjun Lee}, {and} \bibinfo{person}{Hal
  Daum{\'e}~III}.} \bibinfo{year}{2022}\natexlab{}.
\newblock \showarticletitle{What's Different between Visual Question Answering
  for Machine" Understanding" Versus for Accessibility?}
\newblock \bibinfo{journal}{\emph{arXiv preprint arXiv:2210.14966}}
  (\bibinfo{year}{2022}).
\newblock


\bibitem[Chen et~al\mbox{.}(2021)]%
        {chen2021gestonhmd}
\bibfield{author}{\bibinfo{person}{Taizhou Chen}, \bibinfo{person}{Lantian Xu},
  \bibinfo{person}{Xianshan Xu}, {and} \bibinfo{person}{Kening Zhu}.}
  \bibinfo{year}{2021}\natexlab{}.
\newblock \showarticletitle{Gestonhmd: Enabling gesture-based interaction on
  low-cost vr head-mounted display}.
\newblock \bibinfo{journal}{\emph{IEEE Transactions on Visualization and
  Computer Graphics}} \bibinfo{volume}{27}, \bibinfo{number}{5}
  (\bibinfo{year}{2021}), \bibinfo{pages}{2597--2607}.
\newblock


\bibitem[Chiu et~al\mbox{.}(2020)]%
        {chiu2020assessing}
\bibfield{author}{\bibinfo{person}{Tai-Yin Chiu}, \bibinfo{person}{Yinan Zhao},
  {and} \bibinfo{person}{Danna Gurari}.} \bibinfo{year}{2020}\natexlab{}.
\newblock \showarticletitle{Assessing image quality issues for real-world
  problems}. In \bibinfo{booktitle}{\emph{proceedings of the IEEE/CVF
  conference on computer vision and pattern recognition}}.
  \bibinfo{pages}{3646--3656}.
\newblock


\bibitem[{Deng} et~al\mbox{.}(2009)]%
        {deng2009imagenet}
\bibfield{author}{\bibinfo{person}{J. {Deng}}, \bibinfo{person}{W. {Dong}},
  \bibinfo{person}{R. {Socher}}, \bibinfo{person}{L. {Li}},
  \bibinfo{person}{{Kai Li}}, {and} \bibinfo{person}{{Li Fei-Fei}}.}
  \bibinfo{year}{2009}\natexlab{}.
\newblock \showarticletitle{ImageNet: A large-scale hierarchical image
  database}. In \bibinfo{booktitle}{\emph{2009 IEEE Conference on Computer
  Vision and Pattern Recognition}}. \bibinfo{pages}{248--255}.
\newblock


\bibitem[Errattahi et~al\mbox{.}(2018)]%
        {errattahi2018automatic}
\bibfield{author}{\bibinfo{person}{Rahhal Errattahi}, \bibinfo{person}{Asmaa
  El~Hannani}, {and} \bibinfo{person}{Hassan Ouahmane}.}
  \bibinfo{year}{2018}\natexlab{}.
\newblock \showarticletitle{Automatic speech recognition errors detection and
  correction: A review}.
\newblock \bibinfo{journal}{\emph{Procedia Computer Science}}
  \bibinfo{volume}{128} (\bibinfo{year}{2018}), \bibinfo{pages}{32--37}.
\newblock


\bibitem[Gamage et~al\mbox{.}(2023)]%
        {gamage2023what}
\bibfield{author}{\bibinfo{person}{Bhanuka Gamage}, \bibinfo{person}{Thanh-Toan
  Do}, \bibinfo{person}{Nicholas Seow~Chiang Price}, \bibinfo{person}{Arthur
  Lowery}, {and} \bibinfo{person}{Kim Marriott}.}
  \bibinfo{year}{2023}\natexlab{}.
\newblock \showarticletitle{What do Blind and Low-Vision People Really Want
  from Assistive Smart Devices? Comparison of the Literature with a Focus
  Study}. In \bibinfo{booktitle}{\emph{Proceedings of the 25th International
  ACM SIGACCESS Conference on Computers and Accessibility}} (<conf-loc>,
  <city>New York</city>, <state>NY</state>, <country>USA</country>,
  </conf-loc>) \emph{(\bibinfo{series}{ASSETS '23})}.
  \bibinfo{publisher}{Association for Computing Machinery},
  \bibinfo{address}{New York, NY, USA}, Article \bibinfo{articleno}{30},
  \bibinfo{numpages}{21}~pages.
\newblock
\showISBNx{9798400702204}
\urldef\tempurl%
\url{https://doi.org/10.1145/3597638.3608955}
\showDOI{\tempurl}


\bibitem[Ghannay et~al\mbox{.}(2015a)]%
        {ghannay2015asr}
\bibfield{author}{\bibinfo{person}{Sahar Ghannay}, \bibinfo{person}{Nathalie
  Camelin}, {and} \bibinfo{person}{Yannick Esteve}.}
  \bibinfo{year}{2015}\natexlab{a}.
\newblock \showarticletitle{Which ASR errors are hard to detect}. In
  \bibinfo{booktitle}{\emph{Errors by Humans and Machines in Multimedia,
  Multimodal and Multilingual Data Processing (ERRARE 2015) Workshop, Sinaia,
  Romania}}. \bibinfo{pages}{11--13}.
\newblock


\bibitem[Ghannay et~al\mbox{.}(2015b)]%
        {ghannay2015word}
\bibfield{author}{\bibinfo{person}{Sahar Ghannay}, \bibinfo{person}{Yannick
  Esteve}, {and} \bibinfo{person}{Nathalie Camelin}.}
  \bibinfo{year}{2015}\natexlab{b}.
\newblock \showarticletitle{Word embeddings combination and neural networks for
  robustness in asr error detection}. In \bibinfo{booktitle}{\emph{2015 23rd
  European Signal Processing Conference (EUSIPCO)}}. IEEE,
  \bibinfo{pages}{1671--1675}.
\newblock


\bibitem[Goldwater et~al\mbox{.}(2010)]%
        {goldwater2010words}
\bibfield{author}{\bibinfo{person}{Sharon Goldwater}, \bibinfo{person}{Dan
  Jurafsky}, {and} \bibinfo{person}{Christopher~D Manning}.}
  \bibinfo{year}{2010}\natexlab{}.
\newblock \showarticletitle{Which words are hard to recognize? Prosodic,
  lexical, and disfluency factors that increase speech recognition error
  rates}.
\newblock \bibinfo{journal}{\emph{Speech Communication}} \bibinfo{volume}{52},
  \bibinfo{number}{3} (\bibinfo{year}{2010}), \bibinfo{pages}{181--200}.
\newblock


\bibitem[Gonzalez~Penuela et~al\mbox{.}(2024)]%
        {gonzalez2024investigating}
\bibfield{author}{\bibinfo{person}{Ricardo~E Gonzalez~Penuela},
  \bibinfo{person}{Jazmin Collins}, \bibinfo{person}{Cynthia Bennett}, {and}
  \bibinfo{person}{Shiri Azenkot}.} \bibinfo{year}{2024}\natexlab{}.
\newblock \showarticletitle{Investigating Use Cases of AI-Powered Scene
  Description Applications for Blind and Low Vision People}. In
  \bibinfo{booktitle}{\emph{Proceedings of the CHI Conference on Human Factors
  in Computing Systems}}. \bibinfo{pages}{1--21}.
\newblock


\bibitem[Goodfellow et~al\mbox{.}(2014)]%
        {goodfellow2014explaining}
\bibfield{author}{\bibinfo{person}{Ian~J Goodfellow}, \bibinfo{person}{Jonathon
  Shlens}, {and} \bibinfo{person}{Christian Szegedy}.}
  \bibinfo{year}{2014}\natexlab{}.
\newblock \showarticletitle{Explaining and harnessing adversarial examples}.
\newblock \bibinfo{journal}{\emph{arXiv preprint arXiv:1412.6572}}
  (\bibinfo{year}{2014}).
\newblock


\bibitem[Guerreiro et~al\mbox{.}(2018)]%
        {guerreiro2018context}
\bibfield{author}{\bibinfo{person}{Jo{\~a}o Guerreiro}, \bibinfo{person}{Eshed
  Ohn-Bar}, \bibinfo{person}{Dragan Ahmetovic}, \bibinfo{person}{Kris Kitani},
  {and} \bibinfo{person}{Chieko Asakawa}.} \bibinfo{year}{2018}\natexlab{}.
\newblock \showarticletitle{How context and user behavior affect indoor
  navigation assistance for blind people}. In
  \bibinfo{booktitle}{\emph{Proceedings of the 15th International Web for All
  Conference}}. \bibinfo{pages}{1--4}.
\newblock


\bibitem[Gupta et~al\mbox{.}(2020)]%
        {gupta2020hand}
\bibfield{author}{\bibinfo{person}{Sarthak Gupta}, \bibinfo{person}{Siddhant
  Bagga}, {and} \bibinfo{person}{Deepak~Kumar Sharma}.}
  \bibinfo{year}{2020}\natexlab{}.
\newblock \showarticletitle{Hand gesture recognition for human computer
  interaction and its applications in virtual reality}.
\newblock \bibinfo{journal}{\emph{Advanced Computational Intelligence
  Techniques for Virtual Reality in Healthcare}} (\bibinfo{year}{2020}),
  \bibinfo{pages}{85--105}.
\newblock


\bibitem[Herskovitz et~al\mbox{.}(2023)]%
        {herskovitz2023hacking}
\bibfield{author}{\bibinfo{person}{Jaylin Herskovitz}, \bibinfo{person}{Andi
  Xu}, \bibinfo{person}{Rahaf Alharbi}, {and} \bibinfo{person}{Anhong Guo}.}
  \bibinfo{year}{2023}\natexlab{}.
\newblock \showarticletitle{Hacking, switching, combining: understanding and
  supporting DIY assistive technology design by blind people}. In
  \bibinfo{booktitle}{\emph{Proceedings of the 2023 CHI Conference on Human
  Factors in Computing Systems}}. \bibinfo{pages}{1--17}.
\newblock


\bibitem[Hong et~al\mbox{.}(2022)]%
        {hong2022blind}
\bibfield{author}{\bibinfo{person}{Jonggi Hong}, \bibinfo{person}{Jaina
  Gandhi}, \bibinfo{person}{Ernest~Essuah Mensah},
  \bibinfo{person}{Farnaz~Zamiri Zeraati}, \bibinfo{person}{Ebrima Jarjue},
  \bibinfo{person}{Kyungjun Lee}, {and} \bibinfo{person}{Hernisa Kacorri}.}
  \bibinfo{year}{2022}\natexlab{}.
\newblock \showarticletitle{Blind Users Accessing Their Training Images in
  Teachable Object Recognizers}. In \bibinfo{booktitle}{\emph{Proceedings of
  the 24th International ACM SIGACCESS Conference on Computers and
  Accessibility}} (Athens, Greece) \emph{(\bibinfo{series}{ASSETS '22})}.
  \bibinfo{publisher}{Association for Computing Machinery},
  \bibinfo{address}{New York, NY, USA}, Article \bibinfo{articleno}{14},
  \bibinfo{numpages}{18}~pages.
\newblock
\showISBNx{9781450392587}
\urldef\tempurl%
\url{https://doi.org/10.1145/3517428.3544824}
\showDOI{\tempurl}


\bibitem[Hong et~al\mbox{.}(2020a)]%
        {hong2020crowdsourcing}
\bibfield{author}{\bibinfo{person}{Jonggi Hong}, \bibinfo{person}{Kyungjun
  Lee}, \bibinfo{person}{June Xu}, {and} \bibinfo{person}{Hernisa Kacorri}.}
  \bibinfo{year}{2020}\natexlab{a}.
\newblock \showarticletitle{Crowdsourcing the Perception of Machine Teaching}.
  In \bibinfo{booktitle}{\emph{Proceedings of the 2020 CHI Conference on Human
  Factors in Computing Systems}}. \bibinfo{pages}{1--14}.
\newblock


\bibitem[Hong et~al\mbox{.}(2020b)]%
        {hong2020reviewing}
\bibfield{author}{\bibinfo{person}{Jonggi Hong}, \bibinfo{person}{Christine
  Vaing}, \bibinfo{person}{Hernisa Kacorri}, {and} \bibinfo{person}{Leah
  Findlater}.} \bibinfo{year}{2020}\natexlab{b}.
\newblock \showarticletitle{Reviewing Speech Input with Audio: Differences
  between Blind and Sighted Users}.
\newblock \bibinfo{journal}{\emph{ACM Trans. Access. Comput.}}
  \bibinfo{volume}{13}, \bibinfo{number}{1}, Article \bibinfo{articleno}{2}
  (\bibinfo{date}{April} \bibinfo{year}{2020}), \bibinfo{numpages}{28}~pages.
\newblock
\showISSN{1936-7228}
\urldef\tempurl%
\url{https://doi.org/10.1145/3382039}
\showDOI{\tempurl}


\bibitem[Huang et~al\mbox{.}(2021)]%
        {huang2021evaluation}
\bibfield{author}{\bibinfo{person}{Yi-Jheng Huang}, \bibinfo{person}{Kang-Yi
  Liu}, \bibinfo{person}{Suiang-Shyan Lee}, {and} \bibinfo{person}{I-Cheng
  Yeh}.} \bibinfo{year}{2021}\natexlab{}.
\newblock \showarticletitle{Evaluation of a hybrid of hand gesture and
  controller inputs in virtual reality}.
\newblock \bibinfo{journal}{\emph{International Journal of Human--Computer
  Interaction}} \bibinfo{volume}{37}, \bibinfo{number}{2}
  (\bibinfo{year}{2021}), \bibinfo{pages}{169--180}.
\newblock


\bibitem[Jayant et~al\mbox{.}(2011)]%
        {jayant2011supporting}
\bibfield{author}{\bibinfo{person}{Chandrika Jayant}, \bibinfo{person}{Hanjie
  Ji}, \bibinfo{person}{Samuel White}, {and} \bibinfo{person}{Jeffrey~P
  Bigham}.} \bibinfo{year}{2011}\natexlab{}.
\newblock \showarticletitle{Supporting blind photography}. In
  \bibinfo{booktitle}{\emph{The proceedings of the 13th international ACM
  SIGACCESS conference on Computers and accessibility}}.
  \bibinfo{pages}{203--210}.
\newblock


\bibitem[Jiang(2005)]%
        {jiang2005confidence}
\bibfield{author}{\bibinfo{person}{Hui Jiang}.}
  \bibinfo{year}{2005}\natexlab{}.
\newblock \showarticletitle{Confidence measures for speech recognition: A
  survey}.
\newblock \bibinfo{journal}{\emph{Speech communication}} \bibinfo{volume}{45},
  \bibinfo{number}{4} (\bibinfo{year}{2005}), \bibinfo{pages}{455--470}.
\newblock


\bibitem[Jin et~al\mbox{.}(2023)]%
        {jin2023guidelines}
\bibfield{author}{\bibinfo{person}{Weina Jin}, \bibinfo{person}{Xiaoxiao Li},
  \bibinfo{person}{Mostafa Fatehi}, {and} \bibinfo{person}{Ghassan Hamarneh}.}
  \bibinfo{year}{2023}\natexlab{}.
\newblock \showarticletitle{Guidelines and evaluation of clinical explainable
  AI in medical image analysis}.
\newblock \bibinfo{journal}{\emph{Medical Image Analysis}}
  \bibinfo{volume}{84} (\bibinfo{year}{2023}), \bibinfo{pages}{102684}.
\newblock


\bibitem[Kacorri et~al\mbox{.}(2017)]%
        {kacorri2017people}
\bibfield{author}{\bibinfo{person}{Hernisa Kacorri}, \bibinfo{person}{Kris~M.
  Kitani}, \bibinfo{person}{Jeffrey~P. Bigham}, {and} \bibinfo{person}{Chieko
  Asakawa}.} \bibinfo{year}{2017}\natexlab{}.
\newblock \showarticletitle{People with Visual Impairment Training Personal
  Object Recognizers: Feasibility and Challenges}. In
  \bibinfo{booktitle}{\emph{Proceedings of the 2017 CHI Conference on Human
  Factors in Computing Systems}} (Denver, Colorado, USA)
  \emph{(\bibinfo{series}{CHI '17})}. \bibinfo{publisher}{Association for
  Computing Machinery}, \bibinfo{address}{New York, NY, USA},
  \bibinfo{pages}{5839–5849}.
\newblock
\showISBNx{9781450346559}
\urldef\tempurl%
\url{https://doi.org/10.1145/3025453.3025899}
\showDOI{\tempurl}


\bibitem[Karam and Schraefel(2006)]%
        {karam2006investigating}
\bibfield{author}{\bibinfo{person}{Maria Karam} {and} \bibinfo{person}{MC
  Schraefel}.} \bibinfo{year}{2006}\natexlab{}.
\newblock \showarticletitle{Investigating user tolerance for errors in
  vision-enabled gesture-based interactions}. In
  \bibinfo{booktitle}{\emph{Proceedings of the working conference on Advanced
  visual interfaces}}. \bibinfo{pages}{225--232}.
\newblock


\bibitem[Kocielnik et~al\mbox{.}(2019)]%
        {kocielnik2019will}
\bibfield{author}{\bibinfo{person}{Rafal Kocielnik}, \bibinfo{person}{Saleema
  Amershi}, {and} \bibinfo{person}{Paul~N. Bennett}.}
  \bibinfo{year}{2019}\natexlab{}.
\newblock \showarticletitle{Will You Accept an Imperfect AI? Exploring Designs
  for Adjusting End-User Expectations of AI Systems}. In
  \bibinfo{booktitle}{\emph{Proceedings of the 2019 CHI Conference on Human
  Factors in Computing Systems}} (Glasgow, Scotland Uk)
  \emph{(\bibinfo{series}{CHI '19})}. \bibinfo{publisher}{Association for
  Computing Machinery}, \bibinfo{address}{New York, NY, USA},
  \bibinfo{pages}{1–14}.
\newblock
\showISBNx{9781450359702}
\urldef\tempurl%
\url{https://doi.org/10.1145/3290605.3300641}
\showDOI{\tempurl}


\bibitem[Kontogiannis(1999)]%
        {kontogiannis1999user}
\bibfield{author}{\bibinfo{person}{Tom Kontogiannis}.}
  \bibinfo{year}{1999}\natexlab{}.
\newblock \showarticletitle{User strategies in recovering from errors in
  man--machine systems}.
\newblock \bibinfo{journal}{\emph{Safety Science}} \bibinfo{volume}{32},
  \bibinfo{number}{1} (\bibinfo{year}{1999}), \bibinfo{pages}{49--68}.
\newblock


\bibitem[Kontogiannis and Malakis(2009)]%
        {kontogiannis2009proactive}
\bibfield{author}{\bibinfo{person}{Tom Kontogiannis} {and}
  \bibinfo{person}{Stathis Malakis}.} \bibinfo{year}{2009}\natexlab{}.
\newblock \showarticletitle{A proactive approach to human error detection and
  identification in aviation and air traffic control}.
\newblock \bibinfo{journal}{\emph{Safety Science}} \bibinfo{volume}{47},
  \bibinfo{number}{5} (\bibinfo{year}{2009}), \bibinfo{pages}{693 -- 706}.
\newblock
\showISSN{0925-7535}
\urldef\tempurl%
\url{https://doi.org/10.1016/j.ssci.2008.09.007}
\showDOI{\tempurl}


\bibitem[Kurakin et~al\mbox{.}(2016)]%
        {kurakin2016adversarial}
\bibfield{author}{\bibinfo{person}{Alexey Kurakin}, \bibinfo{person}{Ian
  Goodfellow}, \bibinfo{person}{Samy Bengio}, {et~al\mbox{.}}}
  \bibinfo{year}{2016}\natexlab{}.
\newblock \bibinfo{title}{Adversarial examples in the physical world}.
\newblock
\newblock


\bibitem[Kuznetsova et~al\mbox{.}(2020)]%
        {kuznetsova2020open}
\bibfield{author}{\bibinfo{person}{Alina Kuznetsova}, \bibinfo{person}{Hassan
  Rom}, \bibinfo{person}{Neil Alldrin}, \bibinfo{person}{Jasper Uijlings},
  \bibinfo{person}{Ivan Krasin}, \bibinfo{person}{Jordi Pont-Tuset},
  \bibinfo{person}{Shahab Kamali}, \bibinfo{person}{Stefan Popov},
  \bibinfo{person}{Matteo Malloci}, \bibinfo{person}{Alexander Kolesnikov},
  {et~al\mbox{.}}} \bibinfo{year}{2020}\natexlab{}.
\newblock \showarticletitle{The open images dataset v4: Unified image
  classification, object detection, and visual relationship detection at
  scale}.
\newblock \bibinfo{journal}{\emph{International journal of computer vision}}
  \bibinfo{volume}{128}, \bibinfo{number}{7} (\bibinfo{year}{2020}),
  \bibinfo{pages}{1956--1981}.
\newblock


\bibitem[Lafreniere et~al\mbox{.}(2021)]%
        {lafreniere2021false}
\bibfield{author}{\bibinfo{person}{Ben Lafreniere}, \bibinfo{person}{Tanya
  R.~Jonker}, \bibinfo{person}{Stephanie Santosa}, \bibinfo{person}{Mark
  Parent}, \bibinfo{person}{Michael Glueck}, \bibinfo{person}{Tovi Grossman},
  \bibinfo{person}{Hrvoje Benko}, {and} \bibinfo{person}{Daniel Wigdor}.}
  \bibinfo{year}{2021}\natexlab{}.
\newblock \showarticletitle{False positives vs. false negatives: The effects of
  recovery time and cognitive costs on input error preference}. In
  \bibinfo{booktitle}{\emph{The 34th Annual ACM Symposium on User Interface
  Software and Technology}}. \bibinfo{pages}{54--68}.
\newblock


\bibitem[Lee et~al\mbox{.}(2020a)]%
        {lee2020development}
\bibfield{author}{\bibinfo{person}{Chanhwi Lee}, \bibinfo{person}{Jaehan Kim},
  \bibinfo{person}{Seoungbae Cho}, \bibinfo{person}{Jinwoong Kim},
  \bibinfo{person}{Jisang Yoo}, {and} \bibinfo{person}{Soonchul Kwon}.}
  \bibinfo{year}{2020}\natexlab{a}.
\newblock \showarticletitle{Development of real-time hand gesture recognition
  for tabletop holographic display interaction using azure kinect}.
\newblock \bibinfo{journal}{\emph{Sensors}} \bibinfo{volume}{20},
  \bibinfo{number}{16} (\bibinfo{year}{2020}), \bibinfo{pages}{4566}.
\newblock


\bibitem[Lee et~al\mbox{.}(2022a)]%
        {lee2022imageexplorer}
\bibfield{author}{\bibinfo{person}{Jaewook Lee}, \bibinfo{person}{Jaylin
  Herskovitz}, \bibinfo{person}{Yi-Hao Peng}, {and} \bibinfo{person}{Anhong
  Guo}.} \bibinfo{year}{2022}\natexlab{a}.
\newblock \showarticletitle{ImageExplorer: Multi-layered touch exploration to
  encourage skepticism towards imperfect AI-generated image captions}. In
  \bibinfo{booktitle}{\emph{Proceedings of the 2022 CHI Conference on Human
  Factors in Computing Systems}}. \bibinfo{pages}{1--15}.
\newblock


\bibitem[Lee et~al\mbox{.}(2022b)]%
        {lee2022lab}
\bibfield{author}{\bibinfo{person}{Kyungjun Lee}, \bibinfo{person}{Jonggi
  Hong}, \bibinfo{person}{Ebrima Jarjue}, \bibinfo{person}{Ernest~Essuah
  Mensah}, {and} \bibinfo{person}{Hernisa Kacorri}.}
  \bibinfo{year}{2022}\natexlab{b}.
\newblock \showarticletitle{From the lab to people's home: lessons from
  accessing blind participants' interactions via smart glasses in remote
  studies}. In \bibinfo{booktitle}{\emph{Proceedings of the 19th international
  web for all conference}}. \bibinfo{pages}{1--11}.
\newblock


\bibitem[Lee et~al\mbox{.}(2019)]%
        {lee2019revisiting}
\bibfield{author}{\bibinfo{person}{Kyungjun Lee}, \bibinfo{person}{Jonggi
  Hong}, \bibinfo{person}{Simone Pimento}, \bibinfo{person}{Ebrima Jarjue},
  {and} \bibinfo{person}{Hernisa Kacorri}.} \bibinfo{year}{2019}\natexlab{}.
\newblock \showarticletitle{Revisiting blind photography in the context of
  teachable object recognizers}. In \bibinfo{booktitle}{\emph{The 21st
  International ACM SIGACCESS Conference on Computers and Accessibility}}.
  \bibinfo{pages}{83--95}.
\newblock


\bibitem[Lee and Kacorri(2019)]%
        {lee2019hands}
\bibfield{author}{\bibinfo{person}{Kyungjun Lee} {and} \bibinfo{person}{Hernisa
  Kacorri}.} \bibinfo{year}{2019}\natexlab{}.
\newblock \showarticletitle{Hands holding clues for object recognition in
  teachable machines}. In \bibinfo{booktitle}{\emph{Proceedings of the 2019 CHI
  Conference on Human Factors in Computing Systems}}. \bibinfo{pages}{1--12}.
\newblock


\bibitem[Lee et~al\mbox{.}(2020b)]%
        {lee2020pedestrian}
\bibfield{author}{\bibinfo{person}{Kyungjun Lee}, \bibinfo{person}{Daisuke
  Sato}, \bibinfo{person}{Saki Asakawa}, \bibinfo{person}{Hernisa Kacorri},
  {and} \bibinfo{person}{Chieko Asakawa}.} \bibinfo{year}{2020}\natexlab{b}.
\newblock \showarticletitle{Pedestrian detection with wearable cameras for the
  blind: A two-way perspective}. In \bibinfo{booktitle}{\emph{Proceedings of
  the 2020 CHI Conference on Human Factors in Computing Systems}}.
  \bibinfo{pages}{1--12}.
\newblock


\bibitem[Liu et~al\mbox{.}(2023)]%
        {liu2023robot}
\bibfield{author}{\bibinfo{person}{Dewen Liu}, \bibinfo{person}{Changfei Li},
  \bibinfo{person}{Jieqiong Zhang}, {and} \bibinfo{person}{Weidong Huang}.}
  \bibinfo{year}{2023}\natexlab{}.
\newblock \showarticletitle{Robot service failure and recovery: Literature
  review and future directions}.
\newblock \bibinfo{journal}{\emph{International Journal of Advanced Robotic
  Systems}} \bibinfo{volume}{20}, \bibinfo{number}{4} (\bibinfo{year}{2023}),
  \bibinfo{pages}{17298806231191606}.
\newblock


\bibitem[Liu et~al\mbox{.}(2020)]%
        {liu2020deep}
\bibfield{author}{\bibinfo{person}{Li Liu}, \bibinfo{person}{Wanli Ouyang},
  \bibinfo{person}{Xiaogang Wang}, \bibinfo{person}{Paul Fieguth},
  \bibinfo{person}{Jie Chen}, \bibinfo{person}{Xinwang Liu}, {and}
  \bibinfo{person}{Matti Pietik{\"a}inen}.} \bibinfo{year}{2020}\natexlab{}.
\newblock \showarticletitle{Deep learning for generic object detection: A
  survey}.
\newblock \bibinfo{journal}{\emph{International journal of computer vision}}
  \bibinfo{volume}{128}, \bibinfo{number}{2} (\bibinfo{year}{2020}),
  \bibinfo{pages}{261--318}.
\newblock


\bibitem[Lukashova-Sanz et~al\mbox{.}(2023)]%
        {lukashova2023influence}
\bibfield{author}{\bibinfo{person}{Olga Lukashova-Sanz},
  \bibinfo{person}{Martin Dechant}, {and} \bibinfo{person}{Siegfried Wahl}.}
  \bibinfo{year}{2023}\natexlab{}.
\newblock \showarticletitle{The Influence of Disclosing the AI Potential Error
  to the User on the Efficiency of User--AI Collaboration}.
\newblock \bibinfo{journal}{\emph{Applied Sciences}} \bibinfo{volume}{13},
  \bibinfo{number}{6} (\bibinfo{year}{2023}), \bibinfo{pages}{3572}.
\newblock


\bibitem[MacLeod et~al\mbox{.}(2017)]%
        {macleod2017understanding}
\bibfield{author}{\bibinfo{person}{Haley MacLeod}, \bibinfo{person}{Cynthia~L
  Bennett}, \bibinfo{person}{Meredith~Ringel Morris}, {and}
  \bibinfo{person}{Edward Cutrell}.} \bibinfo{year}{2017}\natexlab{}.
\newblock \showarticletitle{Understanding blind people's experiences with
  computer-generated captions of social media images}. In
  \bibinfo{booktitle}{\emph{Proceedings of the 2017 CHI Conference on Human
  Factors in Computing Systems}}. \bibinfo{pages}{5988--5999}.
\newblock


\bibitem[Macrae(2022)]%
        {macrae2022learning}
\bibfield{author}{\bibinfo{person}{Carl Macrae}.}
  \bibinfo{year}{2022}\natexlab{}.
\newblock \showarticletitle{Learning from the failure of autonomous and
  intelligent systems: Accidents, safety, and sociotechnical sources of risk}.
\newblock \bibinfo{journal}{\emph{Risk analysis}} \bibinfo{volume}{42},
  \bibinfo{number}{9} (\bibinfo{year}{2022}), \bibinfo{pages}{1999--2025}.
\newblock


\bibitem[Mandal et~al\mbox{.}(2023)]%
        {mandal2023helping}
\bibfield{author}{\bibinfo{person}{Maniratnam Mandal}, \bibinfo{person}{Deepti
  Ghadiyaram}, \bibinfo{person}{Danna Gurari}, {and} \bibinfo{person}{Alan~C
  Bovik}.} \bibinfo{year}{2023}\natexlab{}.
\newblock \showarticletitle{Helping Visually Impaired People Take Better
  Quality Pictures}.
\newblock \bibinfo{journal}{\emph{IEEE Transactions on Image Processing}}
  (\bibinfo{year}{2023}).
\newblock


\bibitem[Massiceti et~al\mbox{.}(2023)]%
        {massiceti2023explaining}
\bibfield{author}{\bibinfo{person}{Daniela Massiceti}, \bibinfo{person}{Camilla
  Longden}, \bibinfo{person}{Agnieszka Slowik}, \bibinfo{person}{Samuel Wills},
  \bibinfo{person}{Martin Grayson}, {and} \bibinfo{person}{Cecily Morrison}.}
  \bibinfo{year}{2023}\natexlab{}.
\newblock \showarticletitle{Explaining CLIP's performance disparities on data
  from blind/low vision users}.
\newblock \bibinfo{journal}{\emph{arXiv preprint arXiv:2311.17315}}
  (\bibinfo{year}{2023}).
\newblock


\bibitem[Meghana et~al\mbox{.}(2020)]%
        {meghana2020hand}
\bibfield{author}{\bibinfo{person}{M Meghana}, \bibinfo{person}{Ch~Usha
  Kumari}, \bibinfo{person}{J~Sthuthi Priya}, \bibinfo{person}{P Mrinal},
  \bibinfo{person}{K~Abhinav~Venkat Sai}, \bibinfo{person}{S~Prashanth Reddy},
  \bibinfo{person}{K Vikranth}, \bibinfo{person}{T~Santosh Kumar}, {and}
  \bibinfo{person}{Asisa~Kumar Panigrahy}.} \bibinfo{year}{2020}\natexlab{}.
\newblock \showarticletitle{Hand gesture recognition and voice controlled
  robot}.
\newblock \bibinfo{journal}{\emph{Materials Today: Proceedings}}
  \bibinfo{volume}{33} (\bibinfo{year}{2020}), \bibinfo{pages}{4121--4123}.
\newblock


\bibitem[Morris(2020)]%
        {morris2020ai}
\bibfield{author}{\bibinfo{person}{Meredith~Ringel Morris}.}
  \bibinfo{year}{2020}\natexlab{}.
\newblock \showarticletitle{AI and Accessibility}.
\newblock \bibinfo{journal}{\emph{Commun. ACM}} \bibinfo{volume}{63},
  \bibinfo{number}{6} (\bibinfo{year}{2020}), \bibinfo{pages}{35--37}.
\newblock


\bibitem[Myers et~al\mbox{.}(2018)]%
        {myers2018patterns}
\bibfield{author}{\bibinfo{person}{Chelsea Myers}, \bibinfo{person}{Anushay
  Furqan}, \bibinfo{person}{Jessica Nebolsky}, \bibinfo{person}{Karina Caro},
  {and} \bibinfo{person}{Jichen Zhu}.} \bibinfo{year}{2018}\natexlab{}.
\newblock \showarticletitle{Patterns for how users overcome obstacles in voice
  user interfaces}. In \bibinfo{booktitle}{\emph{Proceedings of the 2018 CHI
  conference on human factors in computing systems}}. \bibinfo{pages}{1--7}.
\newblock


\bibitem[Nooruddin et~al\mbox{.}(2020)]%
        {nooruddin2020hgr}
\bibfield{author}{\bibinfo{person}{Nooruddin Nooruddin},
  \bibinfo{person}{Rahool Dembani}, {and} \bibinfo{person}{Nizamuddin Maitlo}.}
  \bibinfo{year}{2020}\natexlab{}.
\newblock \showarticletitle{HGR: Hand-gesture-recognition based text input
  method for AR/VR wearable devices}. In \bibinfo{booktitle}{\emph{2020 IEEE
  international conference on systems, man, and cybernetics (SMC)}}. IEEE,
  \bibinfo{pages}{744--751}.
\newblock


\bibitem[Palmeri and Gauthier(2004)]%
        {palmeri2004visual}
\bibfield{author}{\bibinfo{person}{Thomas~J Palmeri} {and}
  \bibinfo{person}{Isabel Gauthier}.} \bibinfo{year}{2004}\natexlab{}.
\newblock \showarticletitle{Visual object understanding}.
\newblock \bibinfo{journal}{\emph{Nature Reviews Neuroscience}}
  \bibinfo{volume}{5}, \bibinfo{number}{4} (\bibinfo{year}{2004}),
  \bibinfo{pages}{291}.
\newblock
\urldef\tempurl%
\url{https://doi.org/10.1038/nrn1364}
\showDOI{\tempurl}


\bibitem[Pearl(2016)]%
        {pearl2016designing}
\bibfield{author}{\bibinfo{person}{Cathy Pearl}.}
  \bibinfo{year}{2016}\natexlab{}.
\newblock \bibinfo{booktitle}{\emph{Designing voice user interfaces: Principles
  of conversational experiences}}.
\newblock \bibinfo{publisher}{" O'Reilly Media, Inc."}.
\newblock


\bibitem[Perello-March et~al\mbox{.}(2021)]%
        {perello2021driver}
\bibfield{author}{\bibinfo{person}{Jaume~R Perello-March},
  \bibinfo{person}{Christopher~G Burns}, \bibinfo{person}{Roger Woodman},
  \bibinfo{person}{Mark~T Elliott}, {and} \bibinfo{person}{Stewart~A Birrell}.}
  \bibinfo{year}{2021}\natexlab{}.
\newblock \showarticletitle{Driver state monitoring: Manipulating reliability
  expectations in simulated automated driving scenarios}.
\newblock \bibinfo{journal}{\emph{IEEE transactions on intelligent
  transportation systems}} \bibinfo{volume}{23}, \bibinfo{number}{6}
  (\bibinfo{year}{2021}), \bibinfo{pages}{5187--5197}.
\newblock


\bibitem[Pradhan et~al\mbox{.}(2023)]%
        {pradhan2023explainable}
\bibfield{author}{\bibinfo{person}{Biswajeet Pradhan}, \bibinfo{person}{Abhirup
  Dikshit}, \bibinfo{person}{Saro Lee}, {and} \bibinfo{person}{Hyesu Kim}.}
  \bibinfo{year}{2023}\natexlab{}.
\newblock \showarticletitle{An explainable AI (XAI) model for landslide
  susceptibility modeling}.
\newblock \bibinfo{journal}{\emph{Applied Soft Computing}}
  \bibinfo{volume}{142} (\bibinfo{year}{2023}), \bibinfo{pages}{110324}.
\newblock


\bibitem[Raats et~al\mbox{.}(2020)]%
        {raats2020trusting}
\bibfield{author}{\bibinfo{person}{Kaspar Raats}, \bibinfo{person}{Vaike Fors},
  {and} \bibinfo{person}{Sarah Pink}.} \bibinfo{year}{2020}\natexlab{}.
\newblock \showarticletitle{Trusting autonomous vehicles: An interdisciplinary
  approach}.
\newblock \bibinfo{journal}{\emph{Transportation Research Interdisciplinary
  Perspectives}}  \bibinfo{volume}{7} (\bibinfo{year}{2020}),
  \bibinfo{pages}{100201}.
\newblock


\bibitem[Rosen et~al\mbox{.}(2013)]%
        {rosen2013media}
\bibfield{author}{\bibinfo{person}{Larry~D Rosen}, \bibinfo{person}{Kelly
  Whaling}, \bibinfo{person}{L~Mark Carrier}, \bibinfo{person}{Nancy~A
  Cheever}, {and} \bibinfo{person}{Jeffrey Rokkum}.}
  \bibinfo{year}{2013}\natexlab{}.
\newblock \showarticletitle{The media and technology usage and attitudes scale:
  An empirical investigation}.
\newblock \bibinfo{journal}{\emph{Computers in human behavior}}
  \bibinfo{volume}{29}, \bibinfo{number}{6} (\bibinfo{year}{2013}),
  \bibinfo{pages}{2501--2511}.
\newblock


\bibitem[Russakovsky et~al\mbox{.}(2015)]%
        {russakovsky2015imagenet}
\bibfield{author}{\bibinfo{person}{Olga Russakovsky}, \bibinfo{person}{Jia
  Deng}, \bibinfo{person}{Hao Su}, \bibinfo{person}{Jonathan Krause},
  \bibinfo{person}{Sanjeev Satheesh}, \bibinfo{person}{Sean Ma},
  \bibinfo{person}{Zhiheng Huang}, \bibinfo{person}{Andrej Karpathy},
  \bibinfo{person}{Aditya Khosla}, \bibinfo{person}{Michael Bernstein},
  {et~al\mbox{.}}} \bibinfo{year}{2015}\natexlab{}.
\newblock \showarticletitle{Imagenet large scale visual recognition challenge}.
\newblock \bibinfo{journal}{\emph{International Journal of Computer Vision}}
  \bibinfo{volume}{115}, \bibinfo{number}{3} (\bibinfo{year}{2015}),
  \bibinfo{pages}{211--252}.
\newblock


\bibitem[Saha et~al\mbox{.}(2019)]%
        {saha2019closing}
\bibfield{author}{\bibinfo{person}{Manaswi Saha}, \bibinfo{person}{Alexander~J
  Fiannaca}, \bibinfo{person}{Melanie Kneisel}, \bibinfo{person}{Edward
  Cutrell}, {and} \bibinfo{person}{Meredith~Ringel Morris}.}
  \bibinfo{year}{2019}\natexlab{}.
\newblock \showarticletitle{Closing the gap: Designing for the last-few-meters
  wayfinding problem for people with visual impairments}. In
  \bibinfo{booktitle}{\emph{The 21st international acm sigaccess conference on
  computers and accessibility}}. \bibinfo{pages}{222--235}.
\newblock


\bibitem[Salisbury et~al\mbox{.}(2017)]%
        {salisbury2017toward}
\bibfield{author}{\bibinfo{person}{Elliot Salisbury}, \bibinfo{person}{Ece
  Kamar}, {and} \bibinfo{person}{Meredith Morris}.}
  \bibinfo{year}{2017}\natexlab{}.
\newblock \showarticletitle{Toward scalable social alt text: Conversational
  crowdsourcing as a tool for refining vision-to-language technology for the
  blind}. In \bibinfo{booktitle}{\emph{Proceedings of the AAAI Conference on
  Human Computation and Crowdsourcing}}, Vol.~\bibinfo{volume}{5}.
\newblock


\bibitem[Sellen(1994)]%
        {sellen1994detection}
\bibfield{author}{\bibinfo{person}{Abigail~J Sellen}.}
  \bibinfo{year}{1994}\natexlab{}.
\newblock \showarticletitle{Detection of everyday errors}.
\newblock \bibinfo{journal}{\emph{Applied Psychology}} \bibinfo{volume}{43},
  \bibinfo{number}{4} (\bibinfo{year}{1994}), \bibinfo{pages}{475--498}.
\newblock


\bibitem[Sendhilnathan et~al\mbox{.}(2022)]%
        {sendhilnathan2022detecting}
\bibfield{author}{\bibinfo{person}{Naveen Sendhilnathan}, \bibinfo{person}{Ting
  Zhang}, \bibinfo{person}{Ben Lafreniere}, \bibinfo{person}{Tovi Grossman},
  {and} \bibinfo{person}{Tanya~R Jonker}.} \bibinfo{year}{2022}\natexlab{}.
\newblock \showarticletitle{Detecting input recognition errors and user errors
  using gaze dynamics in virtual reality}. In
  \bibinfo{booktitle}{\emph{Proceedings of the 35th Annual ACM Symposium on
  User Interface Software and Technology}}. \bibinfo{pages}{1--19}.
\newblock


\bibitem[Seo and Jung(2021)]%
        {seo2021understanding}
\bibfield{author}{\bibinfo{person}{Woosuk Seo} {and} \bibinfo{person}{Hyunggu
  Jung}.} \bibinfo{year}{2021}\natexlab{}.
\newblock \showarticletitle{Understanding the community of blind or visually
  impaired vloggers on YouTube}.
\newblock \bibinfo{journal}{\emph{Universal Access in the Information Society}}
   \bibinfo{volume}{20} (\bibinfo{year}{2021}), \bibinfo{pages}{31--44}.
\newblock


\bibitem[Shin(2021)]%
        {shin2021effects}
\bibfield{author}{\bibinfo{person}{Donghee Shin}.}
  \bibinfo{year}{2021}\natexlab{}.
\newblock \showarticletitle{The effects of explainability and causability on
  perception, trust, and acceptance: Implications for explainable AI}.
\newblock \bibinfo{journal}{\emph{International Journal of Human-Computer
  Studies}}  \bibinfo{volume}{146} (\bibinfo{year}{2021}),
  \bibinfo{pages}{102551}.
\newblock


\bibitem[Stangl et~al\mbox{.}(2023)]%
        {stangl2023dump}
\bibfield{author}{\bibinfo{person}{Abigale Stangl}, \bibinfo{person}{Emma
  Sadjo}, \bibinfo{person}{Pardis Emami-Naeini}, \bibinfo{person}{Yang Wang},
  \bibinfo{person}{Danna Gurari}, {and} \bibinfo{person}{Leah Findlater}.}
  \bibinfo{year}{2023}\natexlab{}.
\newblock \showarticletitle{“Dump it, Destroy it, Send it to Data Heaven”:
  Blind People’s Expectations for Visual Privacy in Visual Assistance
  Technologies}. In \bibinfo{booktitle}{\emph{Proceedings of the 20th
  International Web for All Conference}}. \bibinfo{pages}{134--147}.
\newblock


\bibitem[{Szegedy} et~al\mbox{.}(2016)]%
        {szegedy2016rethinking}
\bibfield{author}{\bibinfo{person}{C. {Szegedy}}, \bibinfo{person}{V.
  {Vanhoucke}}, \bibinfo{person}{S. {Ioffe}}, \bibinfo{person}{J. {Shlens}},
  {and} \bibinfo{person}{Z. {Wojna}}.} \bibinfo{year}{2016}\natexlab{}.
\newblock \showarticletitle{Rethinking the Inception Architecture for Computer
  Vision}. In \bibinfo{booktitle}{\emph{2016 IEEE Conference on Computer Vision
  and Pattern Recognition (CVPR)}}. \bibinfo{pages}{2818--2826}.
\newblock
\showISSN{1063-6919}
\urldef\tempurl%
\url{https://doi.org/10.1109/CVPR.2016.308}
\showDOI{\tempurl}


\bibitem[Tam et~al\mbox{.}(2014)]%
        {tam2014asr}
\bibfield{author}{\bibinfo{person}{Yik-Cheung Tam}, \bibinfo{person}{Yun Lei},
  \bibinfo{person}{Jing Zheng}, {and} \bibinfo{person}{Wen Wang}.}
  \bibinfo{year}{2014}\natexlab{}.
\newblock \showarticletitle{ASR error detection using recurrent neural network
  language model and complementary ASR}. In \bibinfo{booktitle}{\emph{2014 IEEE
  International Conference on Acoustics, Speech and Signal Processing
  (ICASSP)}}. IEEE, \bibinfo{pages}{2312--2316}.
\newblock


\bibitem[Tan et~al\mbox{.}(2022)]%
        {tan2022self}
\bibfield{author}{\bibinfo{person}{Puchuan Tan}, \bibinfo{person}{Xi Han},
  \bibinfo{person}{Yang Zou}, \bibinfo{person}{Xuecheng Qu},
  \bibinfo{person}{Jiangtao Xue}, \bibinfo{person}{Tong Li},
  \bibinfo{person}{Yiqian Wang}, \bibinfo{person}{Ruizeng Luo},
  \bibinfo{person}{Xi Cui}, \bibinfo{person}{Yuan Xi}, {et~al\mbox{.}}}
  \bibinfo{year}{2022}\natexlab{}.
\newblock \showarticletitle{Self-powered gesture recognition wristband enabled
  by machine learning for full keyboard and multicommand input}.
\newblock \bibinfo{journal}{\emph{Advanced Materials}} \bibinfo{volume}{34},
  \bibinfo{number}{21} (\bibinfo{year}{2022}), \bibinfo{pages}{2200793}.
\newblock


\bibitem[Vorm(2018)]%
        {vorm2018assessing}
\bibfield{author}{\bibinfo{person}{Eric~S Vorm}.}
  \bibinfo{year}{2018}\natexlab{}.
\newblock \showarticletitle{Assessing demand for transparency in intelligent
  systems using machine learning}. In \bibinfo{booktitle}{\emph{2018
  Innovations in Intelligent Systems and Applications (INISTA)}}. IEEE,
  \bibinfo{pages}{1--7}.
\newblock


\bibitem[Wang et~al\mbox{.}(2022)]%
        {wang2022and}
\bibfield{author}{\bibinfo{person}{Junhong Wang}, \bibinfo{person}{Yun Li},
  \bibinfo{person}{Zhaoyu Zhou}, \bibinfo{person}{Chengshun Wang},
  \bibinfo{person}{Yijie Hou}, \bibinfo{person}{Li Zhang},
  \bibinfo{person}{Xiangyang Xue}, \bibinfo{person}{Michael Kamp},
  \bibinfo{person}{Xiaolong~Luke Zhang}, {and} \bibinfo{person}{Siming Chen}.}
  \bibinfo{year}{2022}\natexlab{}.
\newblock \showarticletitle{When, where and how does it fail? a
  spatial-temporal visual analytics approach for interpretable object detection
  in autonomous driving}.
\newblock \bibinfo{journal}{\emph{IEEE Transactions on Visualization and
  Computer Graphics}} \bibinfo{volume}{29}, \bibinfo{number}{12}
  (\bibinfo{year}{2022}), \bibinfo{pages}{5033--5049}.
\newblock


\bibitem[Wu et~al\mbox{.}(2022)]%
        {wu2022fault}
\bibfield{author}{\bibinfo{person}{Chenyun Wu}, \bibinfo{person}{Rabia Sehab},
  \bibinfo{person}{Ahmad Akrad}, {and} \bibinfo{person}{Cristina Morel}.}
  \bibinfo{year}{2022}\natexlab{}.
\newblock \bibinfo{title}{Fault diagnosis methods and Fault tolerant control
  strategies for the electric vehicle powertrains}.
\newblock , \bibinfo{numpages}{4840}~pages.
\newblock


\bibitem[Xie et~al\mbox{.}(2024)]%
        {xie2024bubblecam}
\bibfield{author}{\bibinfo{person}{Jingyi Xie}, \bibinfo{person}{Rui Yu},
  \bibinfo{person}{He Zhang}, \bibinfo{person}{Sooyeon Lee},
  \bibinfo{person}{Syed~Masum Billah}, {and} \bibinfo{person}{John~M Carroll}.}
  \bibinfo{year}{2024}\natexlab{}.
\newblock \showarticletitle{BubbleCam: Engaging Privacy in Remote Sighted
  Assistance}. In \bibinfo{booktitle}{\emph{Proceedings of the CHI Conference
  on Human Factors in Computing Systems}}. \bibinfo{pages}{1--16}.
\newblock


\bibitem[Zhang et~al\mbox{.}(2012)]%
        {zhang2012online}
\bibfield{author}{\bibinfo{person}{Guangxiao Zhang}, \bibinfo{person}{Zhuolin
  Jiang}, {and} \bibinfo{person}{Larry~S Davis}.}
  \bibinfo{year}{2012}\natexlab{}.
\newblock \showarticletitle{Online semi-supervised discriminative dictionary
  learning for sparse representation}. In \bibinfo{booktitle}{\emph{Asian
  conference on computer vision}}. Springer, \bibinfo{pages}{259--273}.
\newblock


\bibitem[Zhang et~al\mbox{.}(2024)]%
        {zhang2024designing}
\bibfield{author}{\bibinfo{person}{Lotus Zhang}, \bibinfo{person}{Abigale
  Stangl}, \bibinfo{person}{Tanusree Sharma}, \bibinfo{person}{Yu-Yun Tseng},
  \bibinfo{person}{Inan Xu}, \bibinfo{person}{Danna Gurari},
  \bibinfo{person}{Yang Wang}, {and} \bibinfo{person}{Leah Findlater}.}
  \bibinfo{year}{2024}\natexlab{}.
\newblock \showarticletitle{Designing Accessible Obfuscation Support for Blind
  Individuals’ Visual Privacy Management}. In
  \bibinfo{booktitle}{\emph{Proceedings of the CHI Conference on Human Factors
  in Computing Systems}}. \bibinfo{pages}{1--19}.
\newblock


\bibitem[Zhang et~al\mbox{.}(2023)]%
        {zhang2023imageally}
\bibfield{author}{\bibinfo{person}{Zhuohao~Jerry Zhang},
  \bibinfo{person}{Smirity Kaushik}, \bibinfo{person}{JooYoung Seo},
  \bibinfo{person}{Haolin Yuan}, \bibinfo{person}{Sauvik Das},
  \bibinfo{person}{Leah Findlater}, \bibinfo{person}{Danna Gurari},
  \bibinfo{person}{Abigale Stangl}, {and} \bibinfo{person}{Yang Wang}.}
  \bibinfo{year}{2023}\natexlab{}.
\newblock \showarticletitle{$\{$ImageAlly$\}$: A $\{$Human-AI$\}$ Hybrid
  Approach to Support Blind People in Detecting and Redacting Private Image
  Content}. In \bibinfo{booktitle}{\emph{Nineteenth Symposium on Usable Privacy
  and Security (SOUPS 2023)}}. \bibinfo{pages}{417--436}.
\newblock


\bibitem[Zhao et~al\mbox{.}(2019)]%
        {zhao2019object}
\bibfield{author}{\bibinfo{person}{Zhong-Qiu Zhao}, \bibinfo{person}{Peng
  Zheng}, \bibinfo{person}{Shou-tao Xu}, {and} \bibinfo{person}{Xindong Wu}.}
  \bibinfo{year}{2019}\natexlab{}.
\newblock \showarticletitle{Object detection with deep learning: A review}.
\newblock \bibinfo{journal}{\emph{IEEE transactions on neural networks and
  learning systems}} \bibinfo{volume}{30}, \bibinfo{number}{11}
  (\bibinfo{year}{2019}), \bibinfo{pages}{3212--3232}.
\newblock


\bibitem[Zou et~al\mbox{.}(2023)]%
        {zou2023object}
\bibfield{author}{\bibinfo{person}{Zhengxia Zou}, \bibinfo{person}{Keyan Chen},
  \bibinfo{person}{Zhenwei Shi}, \bibinfo{person}{Yuhong Guo}, {and}
  \bibinfo{person}{Jieping Ye}.} \bibinfo{year}{2023}\natexlab{}.
\newblock \showarticletitle{Object detection in 20 years: A survey}.
\newblock \bibinfo{journal}{\emph{Proc. IEEE}} \bibinfo{volume}{111},
  \bibinfo{number}{3} (\bibinfo{year}{2023}), \bibinfo{pages}{257--276}.
\newblock


\end{thebibliography}

\appendix

\section{Interview Questions}
\label{appendix_interview_questions}

In this section, you'll find the questions posed to the participants during the user study. If a question has multiple-choice options, they're listed in square brackets after the question.

\subsection{Demographic Information}

\begin{itemize}
    \item What is your age?
    \item What is your gender or gender identity? [woman, man, non-binary]
    \item What is your occupation? 
    \item What is your dominant hand? [left, right]
    \item What phone do you use? Do you use the screen reader (\eg, VoiceOver)?
\end{itemize}

\subsubsection{Visual Impairments}
\begin{itemize}
    \item Do you have visual impairments? [yes, no]
    \item Describe your current level of vision.
    \item For how many years have you had this level of vision ability?
\end{itemize}

\subsubsection{Hearing Impairments}
\begin{itemize}
    \item Do you have hearing impairments? [yes, no]
    \item Describe your current level of hearing ability.
    \item For how many years have you had this level of hearing ability?
\end{itemize}

\subsubsection{Motor Impairments}
\begin{itemize}
    \item Do you have motor impairments? [yes, no]
    \item Describe your current level of motor ability.
    \item For how many years have you had this level of motor ability?
\end{itemize}

\subsection{Technology Experience}
\begin{itemize}
    \item How often do you use a mobile device? [never, once a month, several times a month, once a week, several times a week, once a day, several times a day]
    \item How would you classify your level of familiarity with machine learning? [
        \begin{itemize}
            \item not familiar at all (have never heard of machine learning)
            \item slightly familiar (have heard of it but don't know what it does)
            \item somewhat familiar (I have a broad understanding of what it is and what it does)
            \item extremely familiar (I have extensive knowledge of machine learning)
        \end{itemize}
        ]
\end{itemize}

\subsection{Photo-taking Experience}
\begin{itemize}
    \item How often do you take photos or record a video? [never, once a month, several times a month, once a week, several times a week, once a day, several times a day]
    \item How often do you change the setting of the camera or something in the environment? For example, sitting at the same table, light condition, or using flash. [never, once a month, several times a month, once a week, several times a week, once a day, several times a day]
        \begin{itemize}
            \item (if not “never”) Please describe for what tasks and why.
        \end{itemize}
        
    \item How often do you check if a photo is good after taking it? [never, once a month, several times a month, once a week, several times a week, once a day, several times a day]
        \begin{itemize}
            \item Do you have a strategy for checking a photo?
        \end{itemize}
    
    \item Which of the following do you capture with a camera? (select all that apply) [document, people, landscapes, food, objects, others]
        \begin{itemize}
            \item (for each, ) How often do you capture it with a camera?
        \end{itemize}
    \item On what devices do you interact with a camera like a smartphone, computer, smart glasses, or other devices?
    \item With what applications or tasks do you use a camera? For example, posting on social media, video calls, assistive technologies, etc.
        \begin{itemize}
            \item Why do (or don’t) you use a camera with these applications or tasks?
        \end{itemize}
    \item When you take a photo, how often do you feel confident that it was good? [never, very rarely, rarely, occasionally, very frequently, always]
    \item What challenges do you face when taking photos or broadly manipulating a camera?
\end{itemize}

\subsection{Experience with Image-Based Assistive Tools }
(The following questions are asked for each application from the question above ``With what applications or tasks do you use a camera?'')

\begin{itemize}
    \item How often do you use the app/tool? [never, once a month, several times a month, once a week, several times a week, once a day, several times a day]
    \item How often do you use the app/tool when you don't have access to sighted help? [never, once a month, several times a month, once a week, several times a week, once a day, several times a day]
        \begin{itemize}
            \item Can you provide some examples of when this occurs?
        \end{itemize}
    \item How often would you notice that the app/tool was wrong after the fact? [never, once a month, several times a month, once a week, several times a week, once a day, several times a day]
    \item How often do you encounter misrecognitions when you use the app/tool? [never, very rarely, rarely, occasionally, very frequently, always]
    \item How often do you verify the recognition results when you use the app/tool? [never, very rarely, rarely, occasionally, very frequently, always]
        \begin{itemize}
            \item Why?
        \end{itemize}
    \item I find the app/tool to be useful. [strongly disagree, disagree, neither agree nor disagree, agree, strongly agree]
    \item I care about the misrecognitions of the app/tool? 
        \begin{itemize}
            \item Why?
        \end{itemize}
    \item Are there some situations in which you care about the misrecognitions more than others?
    \item It is challenging to detect the misrecognitions. [strongly disagree, disagree, neither agree nor disagree, agree, strongly agree]
    \item On what devices do you use the app/tool?
    \item For what tasks do you use the app/tool?
    \item What mechanisms do you use to detect the misrecognitions if any?
    \item What kinds of objects do you typically try to recognize with the app/tool?
    \item What is your strategy for taking good photos when using the app/tool?
    \item Do you have a sense of how the app/tool works and how it is able to recognize the object? 
    \item How did you learn to use the app/tool when you first installed it?
    \item Do you like any functions or specific interactions with the app/tool?
    \item Do you dislike any functions or specific interactions with the app/tool?
\end{itemize}

\begin{itemize}
    \item Are you aware of any mobile applications that allow you to personalize them by giving photos of objects or people that you care about?
    \begin{itemize}
        \item (If yes) List them. Which of them have you used before? Can you tell me a bit more about your experience?
    \end{itemize}
\end{itemize}

\subsection{Post-Task Questions}
We asked the following questions to the participants after the error identification task.

\begin{itemize}
    \item It was difficult to identify errors made by the object recognizer. [strongly disagree, disagree, neither agree nor disagree, agree, strongly agree]
        \begin{itemize}
            \item Why? 
        \end{itemize}
    \item How did you know when an object was incorrectly recognized?
\end{itemize}

\end{document}